\documentclass[manuscript]{aastex}
\usepackage{amsmath}
\usepackage{arydshln}
\usepackage{multirow}
\usepackage{lscape}
\usepackage{comment}

\usepackage{grffile}

\newcommand{\cch}{$N=3-2, J=7/2-5/2, F=4-3$ and $3-2$}
\newcommand{\cchshort}{$N=3-2$}
\newcommand{\cs}{$J=5-4$}
\newcommand{\so}{$J_N=6_7-5_6$}

\newcommand{\MF}{HCOOCH$_3$}

\newcommand{\FAD}{H$_2$CO}
\newcommand{\fad}{$5_{1,5}-4_{1,4}$}
\newcommand{\TFA}{H$_2$CS}

\newcommand{\MN}{CH$_3$OH}

\newcommand{\iras}{IRAS 16293--2422}
\newcommand{\irass}{IRAS 15398--3359}
\newcommand{\irasss}{IRAS 18148--0440} 
\newcommand{\irasVLA}{IRAS 05487$+$0255}
\newcommand{\ire}{infalling-rotating envelope}

\newcommand{\rsd}{rotationally-supported disk}
\newcommand{\dfr}{disk-forming region}
\newcommand{\DFR}{Disk-Forming Region}
\newcommand{\cb}{centrifugal barrier} 
\newcommand{\CB}{Centrifugal Barrier}
\newcommand{\centr}{centrifugal radius}
\newcommand{\desys}{disk/envelope system}
\newcommand{\ia}{inclination angle}
\newcommand{\los}{line of sight}

\newcommand{\am}{angular momentum}
\newcommand{\sam}{specific angular momentum}
\newcommand{\SAM}{Specific Angular Momentum}
\newcommand{\dynT}{dynamical time scale} 
\newcommand{\bolL}{bolometric luminosity}
\newcommand{\bolT}{bolometric temperature}
\newcommand{\pos}{plane of the sky}

\newcommand{\Msun}{$M_\odot$}
\newcommand{\Lbol}{$L_{\rm bol}$}
\newcommand{\Lsun}{$L_\odot$}
\newcommand{\Tbol}{$T_{\rm bol}$}
\newcommand{\Mstar}{$M_{\rm star}$}
\newcommand{\Macc}{$\dot{M}_{\rm acc}$}
\newcommand{\Mout}{$\dot{M}_{\rm out}$}
\newcommand{\UnitMrate}{\Msun\ yr\inv}
\newcommand{\rcb}{$r_{\rm CB}$}

\newcommand{\tdyn}{$t_{\rm dyn}$}
\newcommand{\jire}{$j_{\rm IRE}$}

\newcommand{\inv}{$^{-1}$}
\newcommand{\kmps}{km s\inv}

\newcommand{\PAenv}{15\degr}
\newcommand{\PAoutflow}{105\degr}
\newcommand{\vsysval}{5.5}

\newcommand{\parM}{0.15 \Msun}
\newcommand{\parRcb}{100}
\newcommand{\parRcbau}{\parRcb\ au}

\newcommand{\parI}{80\degr}
\newcommand{\incRem}{0\degr\ for a face-on configuration}

\newcommand{\parC}{0.0025 au\inv}
\newcommand{\parV}{0.0015 \kmps}%\ au\inv}

\newcommand{\refOutflowObs}{\citep[e.g.][]{Fuller1995, Hatchell1999, Park2000, Tafalla2000, Jorgensen2004, Takakuwa2007a, Velusamy2014, Leung2016}}
\newcommand{\refOutflowRot}{\citep[e.g.][]{Shu1994a, Shu1994b, Tomisaka2002, Machida2008b, Hartmann2009, Machida2013}}
\newcommand{\refPlots}{\ref{fig:phys_CDt} and \ref{fig:phys_v0Dt}}

\newcommand{\refModels}{Figures \ref{fig:outflow_SEoffset-withmodel} and \ref{fig:outflow_NWoffset-withmodel}}
\newcommand{\refRotzero}{white ellipses}% in \refModels}
\newcommand{\refRotone}{blue ellipses}% in \refModels}
\newcommand{\refRottwice}{green ellipes}% in \refModels}
\newcommand{\refRotonetwice}{blue and green ellipses}% in \refModels}
\newcommand{\refRotforth}{red ellipses}% in \refModels}

\newcounter{tbnotecount}
\newcommand{\tbnote}{\refstepcounter{tbnotecount}\alph{tbnotecount}}
\newcommand{\tbnotemark}{\tablenotemark{\tbnote}}
\newcommand{\tbnotetext}[1]{\tablenotetext{\tbnote}{#1}}
\newcommand{\steptbnote}{\refstepcounter{tbnotecount}}
\newcommand{\resettbnote}[1]{\setcounter{tbnotecount}{0}#1}

\newcommand{\iffigure}{\iftrue}

\renewcommand{\bf}{}

\shorttitle{L483 Outflow}
\shortauthors{Oya et al.}

\title{Sub-arcsecond Kinematic Structure of the Outflow in the Vicinity of the Protostar in L483}
\author{Yoko Oya\altaffilmark{1}, 
Nami Sakai\altaffilmark{2}, 
Yoshimasa Watanabe\altaffilmark{1, 3, 4}, 
Ana L\'{o}pez-Sepulcre\altaffilmark{5, 6, 7}, \\ 
Cecilia Ceccarelli\altaffilmark{6, 7}, 
Bertrand Lefloch\altaffilmark{6, 7}, 
and Satoshi Yamamoto\altaffilmark{1, 8}} 

\email{oya@taurus.phys.s.u-tokyo.ac.jp}
\altaffiltext{1}{Department of Physics, The University of Tokyo, 7-3-1, Hongo, Bunkyo-ku, Tokyo 113-0033, Japan}
\altaffiltext{2}{RIKEN Cluster for Pioneering Research, 2-1, Hirosawa, Wako-shi, Saitama 351-0198, Japan}
\altaffiltext{3}{Division of Physics, Faculty of Pure and Applied Sciences, University of Tsukuba,  Tsukuba, Ibaraki 305-8571, Japan}
\altaffiltext{4}{Tomonaga Center for the History of the Universe, Faculty of Pure and Applied Sciences, University of Tsukuba,  Tsukuba, Ibaraki 305-8571, Japan}
\altaffiltext{5}{Institut de Radioastronomie Millim\'etrique (IRAM), 38406, Saint Martin d'H\'eres, France}
\altaffiltext{6}{Universit\'e Grenoble Alpes, IPAG, F-38000 Grenoble, France}
\altaffiltext{7}{CNRS, IPAG, F-38000 Grenoble, France}
\altaffiltext{8}{Research Center for the Early Universe, The University of Tokyo, 7-3-1, Hongo, Bunkyo-ku, Tokyo 113-0033, Japan} 

\begin{abstract}
The bipolar outflow associated with the Class 0 low-mass protostellar source (IRAS 18148--0440) in L483 has been studied 
in the CCH and CS line emission at 245 and 262 GHz, respectively. 
Sub-arcsecond resolution observations of these lines have been conducted with ALMA. 
Structures and kinematics of the outflow cavity wall are investigated in the CS line, 
and are analyzed by using a parabolic model of an outflow. 
We constrain the inclination angle of the outflow to be from 75\degr\ to 90\degr, 
i.e. the outflow is blowing almost perpendicular to the line of sight. 
Comparing the outflow parameters derived from the model analysis with those of other sources, 
we confirm that the opening angle of the outflow and the gas velocity on its cavity wall 
correlate with the dynamical timescale of the outflows. 
Moreover, a hint of a rotating motion of the outflow cavity wall is found.  
Although the rotation motion is marginal, 
the specific angular momentum of the gas on the outflow cavity wall is evaluated 
to be comparable to or twice that of the infalling-rotating envelope of L483.

\end{abstract}

\keywords{ISM: individual objects (L483, IRAS 18148--0440) -- ISM: molecules -- Stars: formation -- Stars: pre-main -- Stars: outflows}

\begin{document}
\section{Introduction} \label{sec:intro}

\subsection{Outflows in \DFR s} \label{sec:intro_outflow}

Disk formation around newly born solar-type protostars 
has extensively been studied both observationally and theoretically 
%as one of the most important subjects in star-formation studies. 
as one of central issues in astronomy and astrophysics. 
Especially, observational studies are rapidly being developed in the radio astronomy field, 
because high angular-resolution observations down to the disk-forming scale are becoming feasible with the advent of ALMA 
\citep[e.g.,][]{Ohashi_2014_L1527, Sakai_1527nature, Sakai_1527apjl, Sakai_TMC1A, Oya_15398, Oya_1527, Oya_16293, Oya_483, Oya_16293B, Jorgensen_Cycle1, Takakuwa_2017_L1551NE, Aso_2017_L1527, Seifried_Class0disk, Yen_2017_diskgrowth, Alves_BHB07-11, Lee_HH212, Bianchi_2017_HH212}.

In these years, 
%we have revealed 
{ 
the detailed molecular distributions in the \dfr s have been revealed with ALMA %of low-mass protostellar sources 
%in their earliest evolutionary stages (Class 0--I), 
in %low-mass 
protostellar sources in their earliest evolutionary stages (Class 0--I), 
and it has been demonstrated that %a kinematic structure of their \ire\ can be explained by the simple ballistic model 
a gas motion of their \ire s can be interpreted by the simple model assuming a ballistic motion 
\citep[e.g.,][]{Sakai_1527nature, %Sakai_1527apjl, 
Sakai_TMC1A, 
Oya_15398, %Oya_1527, 
Oya_16293, %Oya_483, Oya_16293B, 
Maureia_L1451, Beuther2017_highmass, Alves_BHB07-11, Lee_HH212, Girart2017, vantHoff_L1527, Csengeri2018_SPARKS}. 
}
These results indicate that the gas falls beyond the \centr\ and even to a half of it (`perihelion'). 
%In this model, the energy and angular momentum conservations are assumed, 
%and as a result, the gas cannot fall inward of a certain radius (`perihelion'). 
This position is called `\cb'. %whose radius is twice the \centr. 
More importantly, 
\cb\ is suggested to be a boundary interfacing the infalling envelope with the \rsd\ inside it; 
the gas motion of the \desys\ outside the \cb\ can be regarded as the infaling-rotating motion, 
while that inside it can be as the Keplerian motion. 
These findings opened a new avenue to explore {\it the transition from the envelope to the disk}. 

An important issue to be solved in the disk formation studies is %a mechanism of the loss of the \sam\ of the infalling gas. 
how the \sam\ of the envelope gas is extracted to allow the gas to fall beyond the \cb\ for disk formation. 
At the \cb, the \sam\ of envelope gas is larger than that in the Keplerian disk by a factor of $\sqrt{2}$ (Appendix \ref{sec:appendix_sam}). 
%Thus, the gas has to lose its \sam\ to fall from the \cb\ to the Keplerian disk with some mechanisms. 
%Thus, the \sam\ of the infalling gas has to be extracted by some mechanisms to form the Keplerian disk inside the \cb. 
For the extraction mechanisms, 
outflow launching has been thought to play an important role 
%be a candidate mechanism to extract the \sam\ of the gas in some sources 
\refOutflowRot. 
In fact, rotating motion of outflows/jets has been reported 
{ \citep[e.g.,][]{Zapata_rotjetOrion, Hirota_rotatingOutflow, Lee_rotjetHH212, Alves_BHB07-11}}. 
To elucidate rotating motion of outflows/jets in terms of disk formation, 
observations of both an outflow/jet system and a \desys\ in the vicinity of a protostar is essential. 
We here characterize the molecular outflow of the low-mass protostellar source L483, 
for which the kinematic structure of the \ire\ is already reported \citep{Oya_483}.

\subsection{Target Source: L483} \label{sec:intro_483}
L483 is a dark cloud in Aquila Rift, whose distance from the Sun is 200 pc \citep{Jorgensen2002, Rice2006}. 
%The Class 0 protostar \irasss\  is involved in L483 
This cloud is associated with the infrared source \irasss, which is known to be a Class 0 protostar 
\citep{Fuller1995, Chapman2013}. 
%According to \citet{Shirley2000}, its \bolL\ is 13 \Lsun. 
The bolometric luminosity is 13 \Lsun, according to \citet{Shirley2000}. 
The systemic velocity of this source is 5.5 \kmps\ \citep{Hirota_carbonchain}. 
%The large-scale outflow of L483 has extensively been studied 
Extensive studies have been reported for the large-scale outflow of this source \refOutflowObs. 
%From these studies, the outflow is known to be extended along the east-west direction. 
These studies reveal that the outflow blows along the east-west direction. 
The eastern component is red-shifted, while the western component is blue-shifted. 
\citet{Park2000} reported the position angle of the outflow axis in the \pos\ to be 
95\degr\ on the basis of the HCO$^+$ ($J=1-0$) observation. 
Meanwhile, \citet{Chapman2013} reported it to be 105\degr\ on the basis of the shocked H$_2$ emission \citep{Fuller1995}. 
\citet{Fuller1995} claimed that 
the inclination angle of the outflow is about 50\degr\ with respect to our \los. %(\incRem). 

Recently, we reported the observation of this source in various molecular lines with ALMA \citep{Oya_483}. 
The CS (\cs; 245 GHz) emission was found to be mostly concentrated around the protostar, 
and its kinematic structure was interpreted as %an \ire\ in combination with the Keplerian disk. 
a combination model of the \ire\ with the Keplerian disk.  
With the aid of the simulations assuming the ballistic motion \citep{Oya_15398}, 
the radius of the \cb, which approximately divides these two kinematic components, 
was estimated to be 100 au. 
Inside the \cb, the emission lines of SO, \MN, \MF, and HNCO were detected. 
In addition to the centrally condensed components, 
an extended component is seen for the CS (\cs) and CCH (\cchshort) lines. 
They were found to trace the bipolar outflow components %at a 1000 au scale around the protostar, 
extended over scales of 1000 au from the protostar, 
according to the preliminary analysis \citep{Oya_483}. 
Characterization of the molecular outflow near its launching point is of potential importance 
in relation to the transition from the \ire\ to the Keplerian disk. 
Here, we report detailed analyses of the outflow in this source.

\section{Observations} \label{sec:obs}

%Observations of L483 were conducted with ALMA 
The ALMA observations of L483 were carried out 
during its Cycle 2 operation (12 June 2014). 
The rotational lines of CCH, CS, and SO %in the frequency range from 244 to 264 GHz 
at 262.0, 244.9, and 261.8 GHz, respectively, were observed (ALMA Band 6). 
The line parameters are shown in Table \ref{tb:lines}. 
Other observational details were reported elsewhere \citep{Oya_483}. 

We obtained images by employing the CLEAN procedure. 
The robustness parameter of the Brigg's weighting was set to be 0.5. 
%Self-calibration was employed to obtain better images. 
Self-calibration with the continuum data significantly improved the image quality. 
An image of the 1.2 mm continuum emission was prepared by averaging line-free channels. 
%In preparation of the line images, 
%the continuum component was directly subtracted from the visibilities. % to obtain the molecular line images. 
We subtracted the continuum component directly from the visibilities to prepare the line image. 
Table \ref{tb:lines} lists the synthesized beam sizes for each line. 
We obtained the root-mean-square (rms) noise level of 0.13 mJy beam$^{-1}$ for the continuum image.  
%For the CCH and CS images, 
%we obtained the rms noise level of 8.2 and 7.6 mJy beam$^{-1}$ from the nearby line-free channels, respectively, 
The rms noise level of the CCH and CS line images is evaluated to be 8.2 and 7.6 mJy beam\inv, respectively, 
from nearby line-free channels, 
where the channel width is 61.030 kHz. % <- These rms values are after selfcal. 
The maximum recoverable size of these observations is 3\farcs3 for the CCH and CS lines, 
and 2\farcs0 for the SO line. 
The primary beam correction is not applied. 

\section{Distribution} \label{sec:distribution}

Figure \ref{fig:mom0_extended} shows the moment 0 images of the CCH, CS, and SO lines %CCH (\cch), CS (\cs), and SO (\so) lines 
overlaid on the 1.2 mm dust continuum image. 
The synthesized beam size for the continuum image is $0\farcs46 \times 0\farcs42$ (P.A. 11.\!\!\degr76). 
%The peak position of the continuum emission is 
We determined the peak position of the continuum emission to be 
($\alpha_{2000}$, $\delta_{2000}$) = (18$^{\rm h}$17$^{\rm m}$29\fs947, -04\degr39$^\prime$39\farcs55), % <- The position before self-calibration
%which is determined by means of the two-dimensional Gaussian fit. 
by using the two-dimensional Gaussian fit. 
{ Figures \ref{fig:mom0_extended}(a), (b), and (d)} were originally reported by \citet{Oya_483}. 

The CCH distribution is extended along the northwest-southeast direction. 
It seems to trace a part of the outflow components previously reported \refOutflowObs, 
although it is heavily resolved out. 
The CS distribution is also extended along the northwest-southeast direction. 
{ 
Although the intensity of the extended components appears faint in Figure \ref{fig:mom0_extended}(b) 
because of the wide velocity range for integration ($-2.5 - 13.5$ \kmps), 
it is clearly seen in Figure \ref{fig:mom0_extended}(c) for the narrower velocity range ($3.0 - 5.5$ and $5.5 - 8.0$ \kmps). 
}
Thus, CS also traces the outflow in spite of a heavy resolving-out problem. 
In addition to the outflow, 
a compact component around the continuum peak position is seen in the CS emission, 
which traces the \desys, as mentioned in Section \ref{sec:intro_483} \citep{Oya_483}. 

In contrast to the { CCH and} CS emission, 
the SO emission only traces the compact component around the continuum peak position in this source. 
Although the SO emission often traces the outflow and the outflow shocks 
\citep[e.g.][]{Bachiller_L1157}, 
such a feature is not seen in this sources. 
{ 
The SO emission likely to trace the disk component near the protostar \citep{Oya_483}, 
although it has been reported to trace shocks near the disk edges in some other sources \citep[e.g.,][]{Sakai_1527nature, Lee_HH111}. 
The emitting region of the SO lines would thus be source-dependent.  
}

{ 
Based on these molecular distributions, 
the CCH and CS emission seem to be appropriate for the outflow analysis. 
However, 
the hyperfine structure of the CCH line often makes its velocity structure complicated. 
In fact, the velocity offset of the hyperfine structure ($\sim 2.5$ \kmps; see Table \ref{tb:lines}) 
is comparable to the velocity range of the outflow components in this source (see also Section \ref{sec:outflow_CS}). 
}
%the velocity structure of the CCH line is complicated by the hyperfine structure (Table \ref{tb:lines}). 
{ Therefore,} we here focus on the outflow components traced by CS. 
%We adopt the position angle (P.A.) of \PAoutflow\ for the outflow axis \citep{Chapman2013}, 
In this paper, the position angle (P.A.) of the outflow is assumed to be \PAoutflow\ \citep{Chapman2013}, 
as shown in Figure \ref{fig:mom0_extended}(b).

\section{Outflow Structure Traced by CS} \label{sec:outflow_CS}

{ 
In Figure \ref{fig:mom0_extended}(c), 
%the northwestern and southeastern lobes show both the red- and blue-shifted components. 
red- and blue-shifted components are mutually overlapped with each other. 
This feature is also seen in 
the moment 1 map of the CS line (Figure \ref{fig:mom0_extended}e). 
It suggests that the outflow blows almost on the \pos. 
%against the previously reported \ia\ of 50\degr\ \citep{Fuller1995}. 
}
Figure \ref{fig:PV_CSoutflow}(a) depicts the position-velocity (PV) diagram (P.A. \PAoutflow) prepared along the outflow axis. 
The PV diagram is complicated at first glance, 
and each lobe of the bipolar outflow has both the blue- and red-shifted components. 
Figure \ref{fig:PV_CSoutflow}(b) is the PV diagram %along the line across the southeastern outflow lobe, 
%as indicated by the arrow (P.A. \PAenv) in Figure \ref{fig:mom0_extended}(b). 
along the line indicated by the arrow (P.A. \PAenv) in Figure \ref{fig:mom0_extended}(b), 
i.e. the line just across the southeastern outflow lobe. 
In this PV diagram, an elliptic feature is clearly seen. 
{ 
This feature can be confirmed also in the CCH emission (Figure \ref{fig:PV_CSoutflow}c), 
although the two hyperfine components are nearly overlapped. 
}
Since the extended component is mostly resolved out in interferometric observations, 
this elliptic feature likely corresponds to the outflow cavity wall. 
%Although the components both red- and blue-shifted from the systemic velocity 
%\citep[\vsysval\ \kmps;][]{Hirota_carbonchain} are seen 
Although both the red- and blue-shifted components are visible 
in Figure \ref{fig:PV_CSoutflow}(b), 
the center velocity of the elliptic feature is slightly red-shifted from the 
systemic velocity \citep[\vsysval\ \kmps;][]{Hirota_carbonchain}. 

These observed features can be explained by the geometrical configuration 
shown in Figure \ref{fig:mom0_extended}(f); 
the outflow blows almost on the \pos, 
and the center velocities of the northwestern and southeastern lobes are blue- and red-shifted, respectively. 
It should be noted that the geometrical configuration of the outflow, 
at least in the vicinity of the protostar, 
seems different from the previous report \citep[$i \sim 50$\degr;][]{Fuller1995} (see Section \ref{sec:outflow_model}). 

Such an outflow feature %is quite similar to the case for 
quite resembles that reported for the low-mass Class 0 source 
\irass\ \citep{Oya_15398, Bjerkeli_15398outflow}. 
The kinematic structure of the outflow cavity wall of \irass\ is well explained by a parabolic model of an outflow \citep{Oya_15398}. 
We therefore conduct a similar model analysis for L483. 
We employ the standard outflow model reported by \citet{Lee_outflow}. 
%The details of the model calculation are described 
Further details of the model are presented in Appendix \ref{sec:appendix_model}.

\section{Model Analysis of the Outflow} \label{sec:outflow_model}

%\subsection{Comparison with the Outflow Model} \label{sec:outflow_bestmodel}
In Figure \ref{fig:PV_CSoutflow}, 
the outflow model result is overlaid in white lines on the PV diagrams of CS. 
It seems to well explain the observed feature of the CS line. 
In the model, %the outflow cavity wall is assumed to have a parabolic shape. 
a parabolic shape of the outflow cavity wall is assumed. 
Furthermore, the velocity on the cavity wall is assumed to be proportional to the distance to the protostar. 
%The physical parameters for the model are as follows; 
The model parameters employed here are as follows; 
$i$ = \parI, $C$ = \parC, and $v_0$ = \parV. 
(see Appendix \ref{sec:appendix_model} for their definitions.) %the definition of the physical parameters.) 
{ 
This model is also shown in Figure \ref{fig:mom0_extended}(e). 
%and it seems consistent with the outflow structure traced by the CS line. 
}
With the \ia\ %larger than *\degr\ or less than *\degr, 
less than 75\degr\ or larger than 90\degr, 
the kinematic structure cannot be reproduced by the model. 
{ 
Thus, the outflow axis almost on the \pos\ is confirmed from this kinematical analysis. 
On the contrary, the \ia\ of this source has previously been reported to be 50\degr\ \citep{Fuller1995}. 
This \ia\ is evaluated from the asymmetric brightness of the two lobes in their near-infrared and submillimeter observations, 
and thus this discrepancy would come from the large uncertainty of the previous value. 
} 
On the basis of this model analysis, 
\citet{Oya_483} employed the \ia\ of \parI\ for the kinematic analysis of the \desys.

\begin{figure}
	\begin{center}
	\iffigure
	%\epsscale{1.0}
	%\plotone{PV.CS.5-4.selfcal.vwid41.manucleaned.robust0.5.niter3000.threshold6mJy.notaper.SW0-10asNW0-10asfromContPeakalongPA105deg.PA015.0deg.outflow.color-3-15sigma.contourPreset.NW_D200.I-10G0.5Vas0.3VP1_lee_k7A11_nomodel.eps} 
	\includegraphics[bb = 20 0 2000 1700, scale = 0.27]{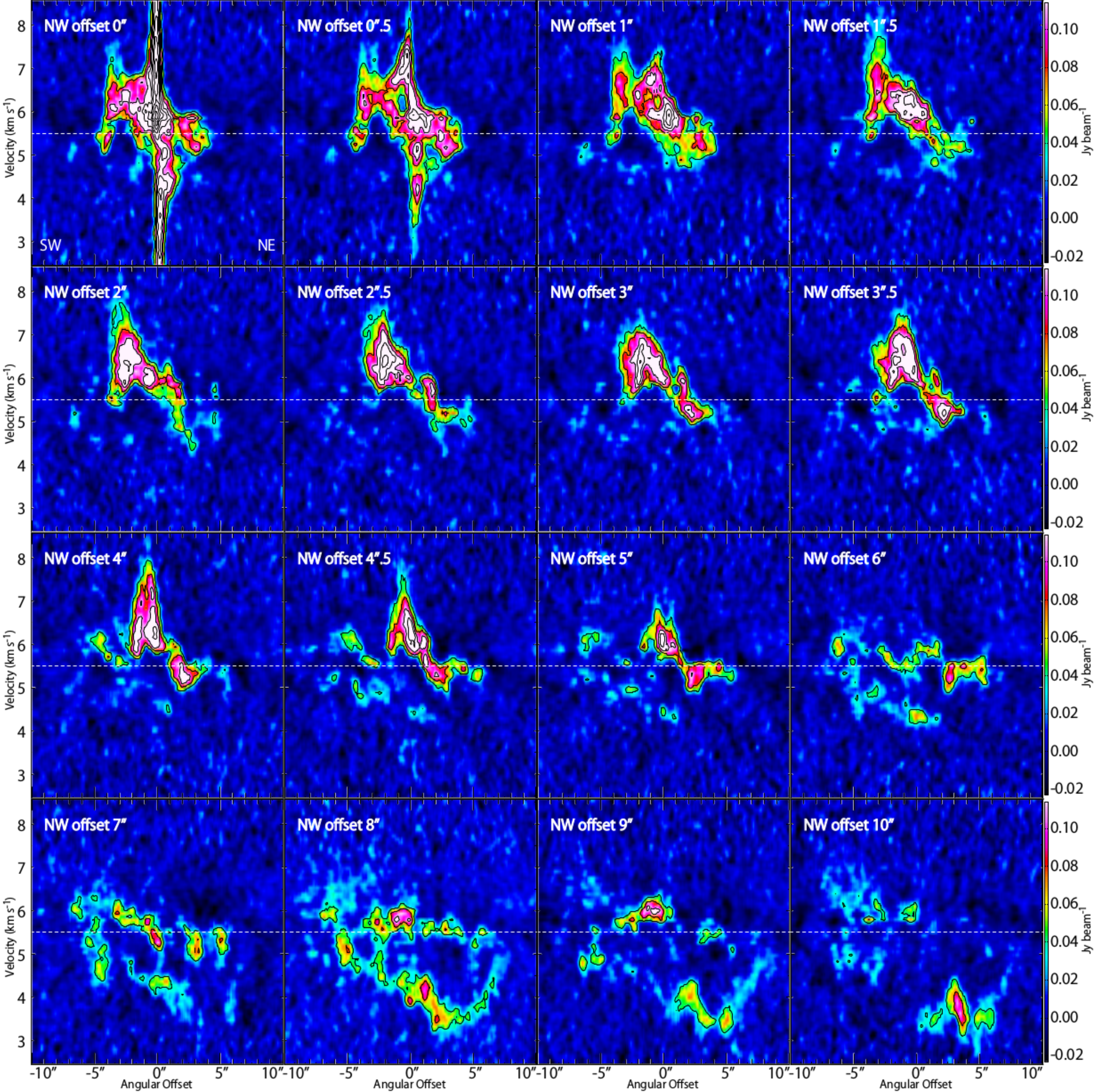}
	\fi
	\caption{Position-velocity diagrams of CS (\cs) across the outflow axis. 
			The position axes are along the position angle of 15\degr, 
			which are perpendicular to the outflow axis (P.A. 105\degr). 
			The origin of the position axes is on the outflow axis with offsets of 
			the distance of ($0\arcsec-10\arcsec$) from the protostellar position 
			toward the northwestern direction. 
			The contour levels for CS are every 5$\sigma$, where the rms noise level is 7.6 mJy beam\inv. 
			\label{fig:outflow_NWoffset-nomodel}}
	\end{center}
\end{figure}

Figure \ref{fig:outflow_SEoffset-nomodel} shows the PV diagrams of CS (\cs), 
%where the position axes are 
which are prepared for the position axes 
taken along the lines with the position angle of \PAenv\ (Figure \ref{fig:mom0_extended}b) 
with the offsets ($0\arcsec-10\arcsec$) from the protostellar position toward the southeastern direction (the red-shifted lobe). 
The origin of each position axis is taken on the outflow axis. 
Figure \ref{fig:outflow_NWoffset-nomodel} is similar to Figure \ref{fig:outflow_SEoffset-nomodel} 
except that the offset is taken toward the northwestern direction (the blue-shifted lobe). 
%The position center of each PV diagram is taken on the outflow axis, and the position axis is taken from southwest to northeast (P.A. \PAenv). 
Although the diagrams seem to be heavily contaminated by the disk/envelope components 
in the panels with the offsets of ($0\arcsec-1\arcsec$) from the protostar, 
the elliptic feature of the outflow component (Figure \ref{fig:PV_CSoutflow}b) can be confirmed in the panels with larger offsets. 
The radial size of the elliptic feature seems to be larger %as the increasing distance from 
for a more distant position to the protostar. 
This feature indicates the expansion of the outflow cavity wall. 
{ 
The velocity centroid is slightly red- and blue-shifted in the southeastern and northwestern lobes, respectively, 
which supports the configuration shown in Figure \ref{fig:mom0_extended}(f). 
The PV diagrams of the two outflow lobes with the same offset from the protostar are not always similar to each other; 
for instance, the diagram with offset of 8\arcsec\ in the blue-shifted lobe (Figure \ref{fig:outflow_NWoffset-nomodel}) 
shows a more extended feature than that in the red-shifted lobe (Figure \ref{fig:outflow_SEoffset-nomodel}). 
This asymmetry would be due to the difference of the ambient environment for the two lobes. 
}

The results of the above outflow model are superposed on these diagrams in { white lines} 
in \refModels. 
In Figure \ref{fig:outflow_NWoffset-withmodel}, 
the observation shows excess of red-shifted velocity in the panels for the angular offset of less than 4\farcs5, 
which possibly comes from a local shock on the outflow cavity wall. 
Such a local shock is also seen in the outflow of \irass\ \citep{Oya_15398}. 
Except for the above shock feature, 
the model results seem to reasonably explain the observed outflow components in all the panels.

By using the physical parameters estimated in the above analysis, 
the \dynT\ (\tdyn) of the outflow lobes is evaluated to be $(3 \pm 1) \times 10^3$ yr with the relation: 
$t_{\rm dyn} = z / v_z = z_0 / v_0$, where $z_0$ is 1 au (see Appendix \ref{sec:appendix_model}). 
\citet{Hatchell1999} previously reported model calculations with the \dynT\ of $(2-6) \times 10^3$ yr 
for the CO ($J=4-3$, $2-1$) observations, 
while \citet{Yildiz_CO-highJ} reported $(4.4 - 6.2) \times 10^3$ yr based on their CO ($J=6-5$, $3-2$) observations. 
On the other hand, \citet{Fuller1995} evaluated it to be $13 \times 10^3$ yr based on the CO ($J=3-2$) observations. 
%they also mentioned that it is inconsistent with the young age of IRAS 18148$-$0440 \citep[$< 4.2 \times 10^3$ yr;][]{Fuller1995}. 
%evaluated from its low \bolT\ \citep[$< 56$ K;][]{Fuller1995}. 
Since they assumed the \ia\ of 50\degr, the \dynT\ is overestimated. 
It is recalculated to be $2 \times 10^3$ yr by use of the \ia\ of 80\degr\ determined in our study. 
Thus, the \dynT s are almost consistent with one another, 
and its most plausible value would be a few $10^3$ yr. 
In particular, it should be stressed that 
the \dynT\ derived from our observations at a 1000 au scale is 
consistent with the above previous reports based on the observations at larger scales 
\citep[e.g. at $\sim 10000$ au scale;][]{Yildiz_CO-highJ}. 
The outflow parameters for this source are summarized in Table \ref{tb:params_483}. 

\section{Evolution of Outflows} \label{sec:phys_outflow}

\subsection{Comparison with Other Sources} \label{sec:phys_outflow_others}
\citet{Oya_15398, Oya_1527} 
reported the kinematic structure of the outflow near the protostar 
for \irass\ and L1527 observed with ALMA 
{ as well as their envelope structure.} 
The physical parameters for the outflow evaluated for these sources with the aid of the parabolic outflow model 
are summarized in Table \ref{tb:phys_outflow}. 

Both \irass\ and L1527 have outflows blowing almost on the \pos, 
as L483. 
However, their outflow shapes are quite different from each other. 
The outflow of \irass\ is well collimated, 
while that of L1527 shows a butterfly-feature. 
The L483 case seems in between. 
In the L1527 case, an offset of 1\farcs24 (170 au) between the launching points of 
blue-shifted and red-shifted lobes is assumed to account for the outflow structure, 
as reported by \citet{Tobin2008}. 
In contrast, such an offset of the outflow launching points is not definitively seen in L483 
as well as \irass. 
This is probably because the envelope component would be contaminated with the outflow component 
near the protostar for the L483 and \irass\ cases. 
%Apart from the offset, 
If such an offset is ignored for simplicity, 
the diversity of the opening angles of the outflow cavity is mainly translated to the variation of $C$ 
by an order of magnitude among these three sources. 

In Table \ref{tb:phys_outflow}, the error ranges of the physical parameters for L483, \irass, and L1527 
are estimated on the basis of the simulations with a wide range of the parameters. 
{ 
The \dynT\ of the outflow is evaluated to be 
$(1.9 \pm 0.2) \times 10^3$ and $(6.5 \pm 1.3) \times 10^3$ yr 
for \irass\ and L1527, respectively. 
These values are comparable to or different by a factor of a few from  
those reported based on larger scale observations 
\citep[Table \ref{tb:phys_outflow};][]{Yildiz_CO-highJ}. 
}
{ 
It would be more appropriate to employ the \dynT s by \citet{Yildiz_CO-highJ} 
rather than those derived from the outflow model. 
}
%since large scale structures are resolved out in the ALMA observations toward these sources. 
%it is not appropriate to estimate the \dynT s of the outflow with these data. 
%Therefore, the \dynT s reported for L1527 and \irass\ %in Table \ref{tb:phys_outflow} are taken from \citet{Yildiz_CO-highJ}. 
%by \citet{Yildiz_CO-highJ} are employed, as listed in Table \ref{tb:phys_outflow}. 
As for the L483 case, 
we employ the \dynT\ estimated in our analysis ($t_{\rm dyn} = 3 \times 10^3$ yr), 
because it seems to be reliable, as discussed in Section \ref{sec:outflow_model}. 

\subsection{Relation to Dynamical Ages} \label{sec:phys_outflow_others} 

We compare the physical parameters of the outflow model ($C$ and $v_0$) and the \dynT\ for the seven sources listed in Table \ref{tb:phys_outflow}. 
{ We also compare the model parameters and the \bolT.} %except for \irasVLA.} 
Figures \refPlots\ show semi-log plots of $C$ and $v_0$ versus the \dynT\ { and \bolT,} respectively. 
{ The plots with the \dynT\ were}
previously reported by \citet{Oya_1527} for the six sources except for L483. 
%The outflow parameters ($C$ and $v_0$) for the seven sources are also summarized in Table \ref{tb:phys_outflow}. 
The outflow parameters for \irasVLA, RNO 91, L1448C, and HH 46/47 are 
converted to those in the units used for the other sources, 
as described in Appendix \ref{sec:appendix_model} \citep[see also][]{Oya_1527}. 
{ 
We excluded \irasVLA\ in Figures \refPlots, 
because the \bolT\ is not available for this source. 
}

As shown in Figures \refPlots, 
the results for L483 seem to be consistent with those for the other sources. 
The dashed lines are obtained by linear fitting to the data under the assumption of equal weight. 
The correlation coefficients are 
{ 
$-0.95$, $-0.85$, %-0.9423604128318065 without L483 -> -0.9489019899058226 with L483 tdyn = 2x10^3 yr -> -0.9470677759909334 with L483 tdyn = 3x10^3 yr
$-0.75$, and $-0.63$ %-0.8022890471843741 without L483 -> -0.7396279581940395 with L483 tdyn = 2x10^3 yr -> -0.7542398314987678 with L483 tdyn = 3x10^3 yr
for the four plots. 
}
%for Figures \refPlots, respectively. 
In spite of a small number of the sources, 
{ we can see a trend that} 
both $C$ and $v_0$ decrease exponentially as an increasing \dynT\ of the outflow { and an increasing \bolT.} 
%which is consistent with the results previously reported by \citet{Oya_1527}. 
%Such a relationship between the opening angle of outflow and the source age is previously reported 
%\citep[e.g.][]{Arce2006, Seale2008, Velusamy2014}. 

\begin{figure}
	\begin{center}
	\iffigure
	%\epsscale{1.0}
	%\plotone{Cvs_tdyn-Tbol.eps}
	%\plotone{c-tdyn.eps}
	\includegraphics[bb = 0 0 1800 100, scale = 0.325]{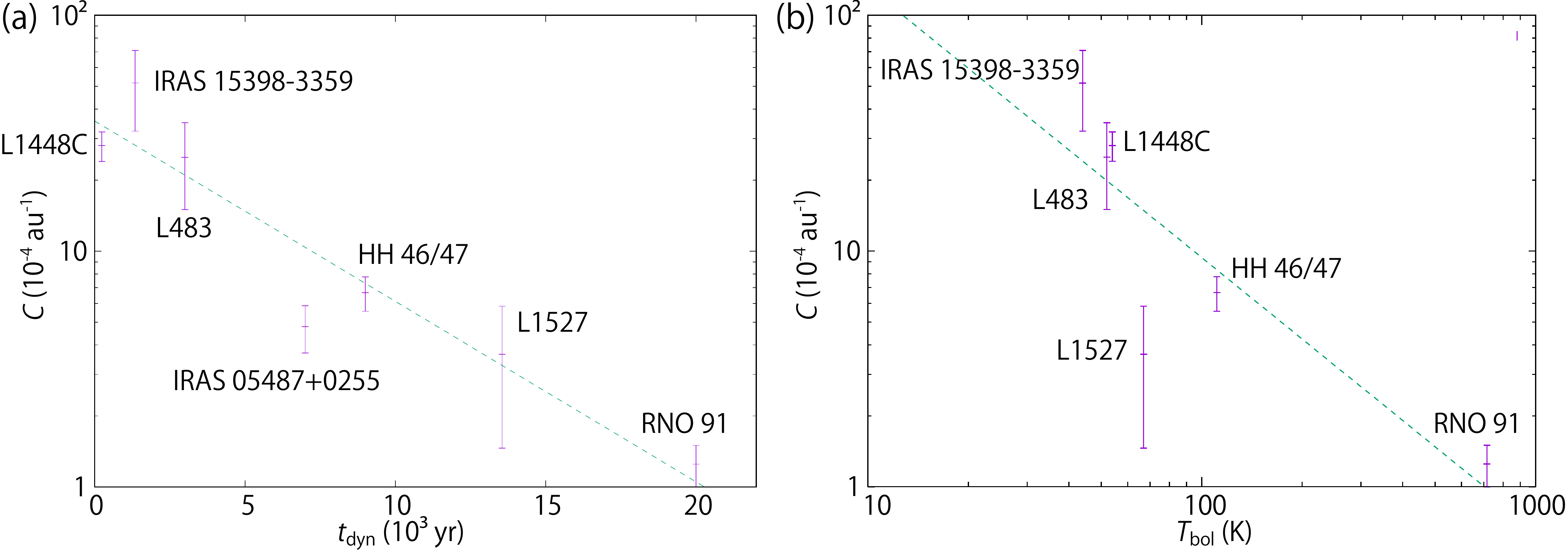}
	\fi
	\caption{Relation between the curvature parameter ($C$) of an outflow, { its \dynT\ (\tdyn; a), and \bolT\ (\Tbol; b).} 
			The \dynT s for \irass\ and L1527 are the averaged value of the blue and red lobes. 
			The green dashed lines represent the best-fit functions: 
			$\log C = (-0.18 \pm 0.03) \times (t_{\rm dyn} \times 10^{-3}) + (-5.6 \pm 0.3)$ %with L483 tdyn = 3x10^3 yr 
			%$\log C = (-0.17 \pm 0.03) \times (t_{\rm dyn} \times 10^{-3}) + (-5.7 \pm 0.3)$, %with L483 tdyn = 2x10^3 yr (almost the same as the case without L483)
			%$\log (C/D) = (-0.17 \pm 0.03) \times (t_{\rm dyn} \times 10^{-3}) + (-5.7 \pm 0.3)$, %without L483
			{ 
			and 
			$\log C = (-1.1 \pm 0.4) \times \log$ \Tbol $ + (-2 \pm 2)$ 
			for (a) and (b), 
			respectively. 
			}
			The correlation coefficients are 
			{ 
			$-0.95$ %$-0.94$. 
			and $-0.85$ 
			 for (a) and (b), respectively, 
			 }
			where the uniform weights are applied to all the sources. 
			\label{fig:phys_CDt}}
	\end{center}
\end{figure}

\begin{figure}
	\begin{center}
	\iffigure
	%\epsscale{0.8}
	%plotone{OpeningAnglevsYSOage.eps}
	\includegraphics[bb = 0 0 1800 500, scale = 0.68]{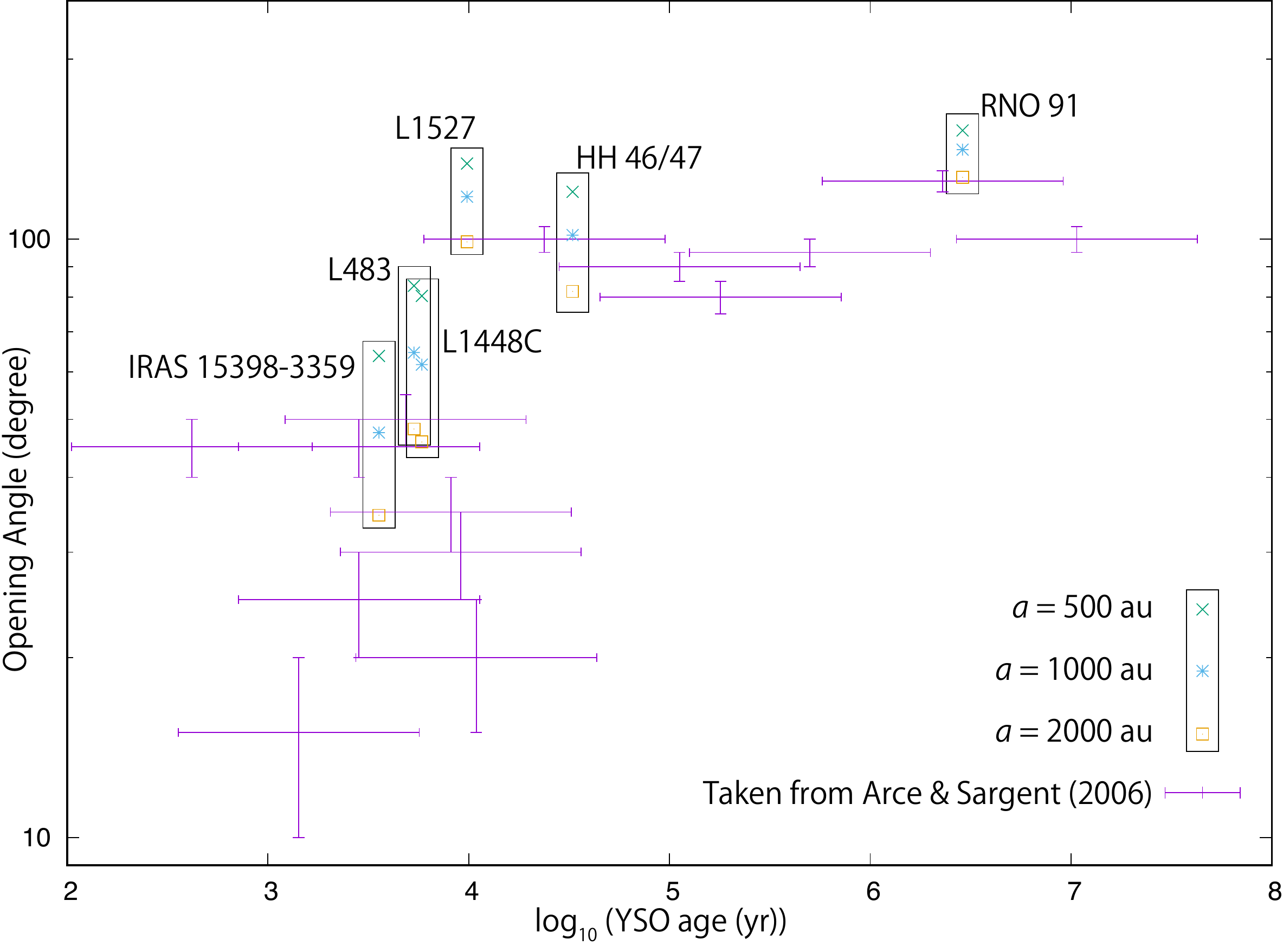}
	\fi
	\caption{ 
			Relation between the opening angle of the outflow and the source age. 
			The original plot was reported by \citet{Arce2006}, 
			from which three sources (L1527, L1448, and RNO91) are replaced 
			by the results of this study.  
			The sources listed in Table \ref{tb:phys_outflow} are also shown in the plot, 
			except for \irasVLA. 
			The source age is obtained from the \bolT\ by using the relation \citep{Ladd1998}: 
			$\log (t_{\rm years}) = [2.4 \times \log (T_{\rm bol}) - 0.9] \pm 0.6$. 
			For the samples taken from \citet{Arce2006}, 
			the errors for the source age come from this relation, 
			while those for the opening angle are uniformly set to be 5\degr. 
			For the other six sources, 
			the opening angles are derived by adopting $a$ of 500, 1000, and 2000 au 
			based on the outflow model analysis, 
			which are approximate extents of the outflows 
			(see Appendix \ref{sec:appendix_model}). 
			\label{fig:opangle}}
	\end{center}
\end{figure}

As discussed by \citet{Oya_1527}, 
the trend seen in Figure \ref{fig:phys_CDt} well corresponds to the previous observational and theoretical results 
which report a relation between an opening angle of an outflow and a source age 
\citep[e.g.][]{Arce2006, Shang2006, Seale2008, Offner2011, Machida2013, Velusamy2014}. 
{ 
Figure \ref{fig:opangle} shows the relation between the opening angle of the outflow and the source age. 
This plot was originally reported by \citet{Arce2006}. 
In order to involve the sources listed in Table \ref{tb:phys_outflow} in this plot, 
we calculated the opening angle of the outflow based on the model results (see Appendix \ref{sec:appendix_model}). 
These new samples are consistent with the trend reported by \citet{Arce2006}. 
}

Again, it should be stressed that we evaluate the outflow parameters in Table \ref{tb:phys_outflow} 
from the full use of the geometrical and velocity structures of the outflow near the protostar 
considering the \ia, and provide quantitative supports of the above trends. 
Moreover, the ALMA results focused on narrow regions (at $\sim$ 1000 au scale; L483, \irass, and L1527) 
are consistent with the other results based on the observations at larger scales. 
Thus, outflows can be characterized by focusing only on a small region around the protostar. 
%without investigating its whole structure. 
%even if its whole structure is not observed. 
This result further supports the idea 
that the outflow launching is mostly defined in the vicinity of the protostar.

\section{Rotation Motion in the Outflow} \label{sec:outflow_rotation}

In protostellar evolution, especially in disk formation, 
\am\ of the gas is expected to play a crucial role. 
The infalling envelope gas needs to lose its \am\ to fall becyond the \cb\ for disk formation %form a \rsd\ 
(Section \ref{sec:intro_outflow}). 
Outflow launching is thought to be a potential mechanism 
%to extract the \am\ of the envelope gas 
for the \am\ extraction 
\refOutflowRot. %at the \cb s. 
If this is the case, outflows would have a rotation motion. 
On the basis of this prediction, we examined the rotation motion of the outflow cavity wall in L1527 \citep{Oya_1527} 
without success because of the insufficient signal-to-noise (S/N) ratio of the available data. 
Here, we conduct such an analysis for L483, where the S/N ratio of the outflow image in the CS line 
is better than that in the L1527 case. 

{ 
In Figures \ref{fig:mom0_extended}(c) and (e), 
we see a hint that there is a velocity gradient perpendicular to the outflow axis, 
although these maps possibly suffered by an asymmetric distribution of the gas or local shocks on the cavity wall. 
Moreover, 
}
when we carefully look at Figure \ref{fig:outflow_SEoffset-withmodel}, 
we may notice that the observed elliptic shape of the PV diagram tends to be elongated obliquely 
for the small offsets (1\farcs5--3\arcsec) in comparison with the simulation { shown in white ellipses.} 
A similar trend could be seen in Figure \ref{fig:outflow_NWoffset-withmodel}. 
Such a slant distortion of the expanding motion in the PV diagram suggests 
association of the rotation motion. 
Although the above trend is marginal, 
we examine the effect of the rotation of the outflow on the PV diagrams. 
The blue, green, and red elliptic lines in Figures \ref{fig:outflow_SEoffset-withmodel} and \ref{fig:outflow_NWoffset-withmodel} 
show the model results superposed on the PV diagrams of CS, 
where the rotation motion of the outflow is considered. 
The rotation velocity of the outflow cavity wall is simply calculated under assumption of the \am\ conservation. 
In the model calculations, 
the \sam\ of the outflowing gas is assumed to be 
the same as that of the \ire\ component \citep[\jire\ = $7.9 \times 10^{-4}$ \kmps\ pc; ][]{Oya_483}
{ 
(\refRotone\ in \refModels), 
%which can be evaluated to be $7.9 \times 10^{-4}$ \kmps\ pc according to \citet{Oya_483}, 
twice as much ($j = 2 \times$ \jire; \refRottwice), or four times as much ($j = 4 \times$ \jire; \refRotforth), 
}
as examples. 
When a larger \sam\ is assumed, a slant distortion of the elliptic feature in the PV diagram becomes more significant. 
Comparing the model results with the observations, 
we find a hint that the first and second models 
{ ($j = $ \jire, $2 \times$ \jire; \refRotonetwice)} 
would better reproduce the observations 
in the panels with the smaller offsets (e.g. an offset of $1\arcsec-4\arcsec$) 
than the model without rotation motion { ($j = 0$; \refRotzero).} 
The rotation motion seems to be overestimated in the third model { ($j = 4 \times$ \jire; \refRotforth).} 
%Thus, the specific angular momentum of the outflowing gas can be assessed to be a factor of few of that of the \ire. 

If the outflow is really extracting a substantial amount of the \am\ from the envelope gas, %at the \cb, 
the outflow should extract a larger \sam\ than that of the envelope. 
{ 
Suppose that a gas clump with the mass of $(m_1 + m_2)$ and the \sam\ of $j_0$ 
splits into two smaller gas clumps of the mass of $m_1$ and $m_2$ 
with the \sam\ of $j_1$ and $j_2$, respectively. 
Then, the conservation of \am\ is expressed as: 
\begin{align}
	(m_1 + m_2) \times j_0 &= m_1 \times j_1 + m_2 \times j_2, \label{eq:AMcons}
\end{align}
If the small gas clump with the mass of $m_1$ falls toward the protostar because of \am\ loss ($j_1 < j_0$), 
the other small gas clump with the mass of $m_2$ is required to extract a larger \sam\ before the split ($j_2 > j_0$). 
}
%Thus, the better match for the second case than the first case seems reasonable. 
%{ 
Thus, a better agreement for the second case { ($j = 2 \times$ \jire; \refRottwice)} 
seems reasonable.
%} 
Moreover, the third { ($j = 4 \times$ \jire; \refRotforth)} case gives the upper limit to the \sam\ that the outflow is extracting. 
{ 
Nevertheless, 
it is difficult to evaluate the amount of 
the \sam\ and the total angular momentum extracted by the outflow accurately at the current stage. 
}

As shown here, 
a change in the \sam\ by a factor of a few can affect the outflow feature in the PV diagrams. 
%feasibility of the detection of the rotation motion in the outflow. 
%We thus need to investigate the kinematic structure of the outflow much more carefully, 
We can thus investigate the outflow rotation by looking at the kinematic structure much more carefully 
with better quality and higher angular resolution data, 
especially in the vicinity of its launching point, 
where the rotation motion is expected to be most prominent. 

It should be noted that \citet{Hirota_rotatingOutflow} recently reported a clear rotation motion of the outflow 
in the high-mass young stellar object candidate Orion Source I. 
They observed the outflow in the SiO line at an angular resolution of $\sim 0\farcs1$ with ALMA, 
and detected rotation velocities of 17.9 and 7.0 \kmps\ at the distances of 24 and 76 au from the protostar, respectively. 
This rotation velocity is translated to the \sam\ of $2.1 \times 10^{-3}$ and $2.6 \times 10^{-3}$ \kmps\ pc, 
which is larger than that of L483 ($7.9 \times 10^{-4}$ \kmps\ pc; Table \ref{tb:params_483}). 

\section{Relation between the Envelope and the Outflow} \label{sec:env-outflow}

\subsection{Mass Rates of Outflow and Accretion} \label{sec:env-outflow_mass}
The outflow mass loss rate of L483 was reported to be $(5.4-14) \times 10^{-6}$ \UnitMrate\ by \citet{Yildiz_CO-highJ}. 
{ Here, we compare this with the accretion rate.} 
The averaged mass accretion rate (\Macc) can be estimated to be $5 \times 10^{-5}$ \UnitMrate\ 
from the \dynT\ of $3 \times 10^3$ yr (Section \ref{sec:outflow_model}) and 
the protostellar mass \citep[\Mstar\ = 0.15 \Msun;][]{Oya_483}. 
%{ 
%The \dynT\ of the outflow would be the lower limit for the actual source age. 
%Therefore, we also roughly estimate the source age to be $(1.3-21) \times 10^3$ yr 
%with the relation between the \bolT\ and the source age reported by \citet{Ladd1998}: 
%$\log (t_{\rm years}) = [2.4 \times \log (T_{\rm bol}) - 0.9] \pm 0.6$. 
%Then, the averaged mass accretion rate is estimated to be $(7 \times 10^{-6} - 1 \times 10^{-4})$ yr. 
%}
%
The current mass accretion rate can also be derived from the \bolL\ by using the following equation \citep{Palla1991}; 
%\citep[\Lbol\ = 13 \Lsun;][]{Shirley2000} 
\begin{align}
	\dot{M}_{\rm acc} &= \frac{L_{\rm bol} R_{\rm star}}{G M_{\rm star}}, \label{eq:Macc}
\end{align}
where $R_{\rm star}$ denotes the protostar radius and $G$ the gravitational constant. 
Assuming $R_{\rm star}$ of 2.5 $R_\odot$ \citep[e.g.][]{Baraffe2010}, 
\Macc\ is calculated to be $6.9 \times 10^{-6}$ \UnitMrate\ by using the current \bolL. 
%{ 
%Although this value seems to be lower than the value derived from the \dynT\ of the outflow, 
%it is consistent with that from the estimated source age. 
%
%Also, 
%}
{ 
Compared with this value, 
the mass accretion rate derived from the \dynT\ of the outflow %($5 \times 10^{-5}$ \UnitMrate) 
may be larger than the outflow mass loss rate. 
Since the \dynT\ is expected to give the lower limit for the source age, 
the mass accretion rate derived from it could be overestimated. 
}
%All these mass accretion rates are comparable with the canonical value 
%\citep[$10^{-6}-10^{-5}$ \UnitMrate;][]{Dunham_massRate}, 
%and are comparable with or larger than the mass-loss rate by the outflow. 

%\subsection{Angular Momentum Loss of Envelope Gas} \label{sec:env-outflow_amloss} 
\subsection{Is the \CB\ the Launching Point of the Outflow?} \label{sec:env-outflow_cb}

Recent studies reveal that the gas motion in the envelope 
at a scale of a few 100 au from the protostar can be explained by the \ire\ model 
\citep{Sakai_1527nature, Sakai_1527apjl, Sakai_TMC1A, Oya_15398, Oya_1527, Oya_16293, Oya_483}. 
Moreover, the disk component seems to exist inside the \cb, 
which can be traced by the \TFA\ and \FAD\ lines 
in \iras\ \citep{Oya_16293} and L1527 \citep{Sakai_1527apjl}, respectively. 
Thus, the \cb\ can be recognized as a boundary interfacing the infalling envelope with the disk component. %\rsd. 
However, the gas of the infalling envelope cannot fall into the disk beyond the \cb, 
as long as its \sam\ is conserved, 
as mentioned before (Section \ref{sec:intro_outflow}). 
To extract the \sam\ of the envelope gas for disk formation and further growth of the protostar, 
the outflow launched around the \cb\ could play an important role. 
A hint of the outflow rotation described in Section \ref{sec:outflow_rotation} may support this idea, if confirmed. 

\citet{Alves_BHB07-11} recently reported a support for this picture; 
the molecular outflow in BHB07-11 is clearly delineated with their ALMA observation, 
and it is found to be launched outside the disk, far from the protostellar position. 
We also found a similar feature in \iras\ Source B; 
the pole-on outflow lobes traced by the SiO line show a radial offset near the protostar, 
and their launching point seems to be near the \cb\ traced by OCS and \TFA\ \citep[\rcb\ $\sim 40$ au;][]{Oya_16293B}. 
In L483, 
the SiO emission shows an intensity peak at the position apart from the protostar by 100 au, 
which is close to the position of the \cb\ \citep{Oya_483}. 
Since SiO is known as a shock tracer \citep[e.g.][]{Mikami_L1157-SiO}, 
the results for \iras\ Source B and L483 imply possible shocks caused by the collision of the outflowing gas 
with the infalling envelope gas near the \cb. 

Recently, \citet{Sakai_1527_highres} reported that the envelope gas in L1527 is accumulated in front of the \cb, 
and that it has a substantial extension perpendicular to the mid-plane. 
It is likely that a part of the gas is escaping from the mid-plane. 
\citet{Sakai_1527_highres} suggested a possibility that this outflowing motion forms 
so-called `disk winds' or `low-velocity molecular outflow' launched at the \cb. 
If so, one would expect a rotation motion of the outflow particularly near the \cb. 

\begin{figure}
	\begin{center}
	\iffigure
	%\epsscale{0.53}
	%\plotone{c-j_cerror_15398smallMass.eps}
	\includegraphics[bb = 0 0 750 100, scale = 0.4]{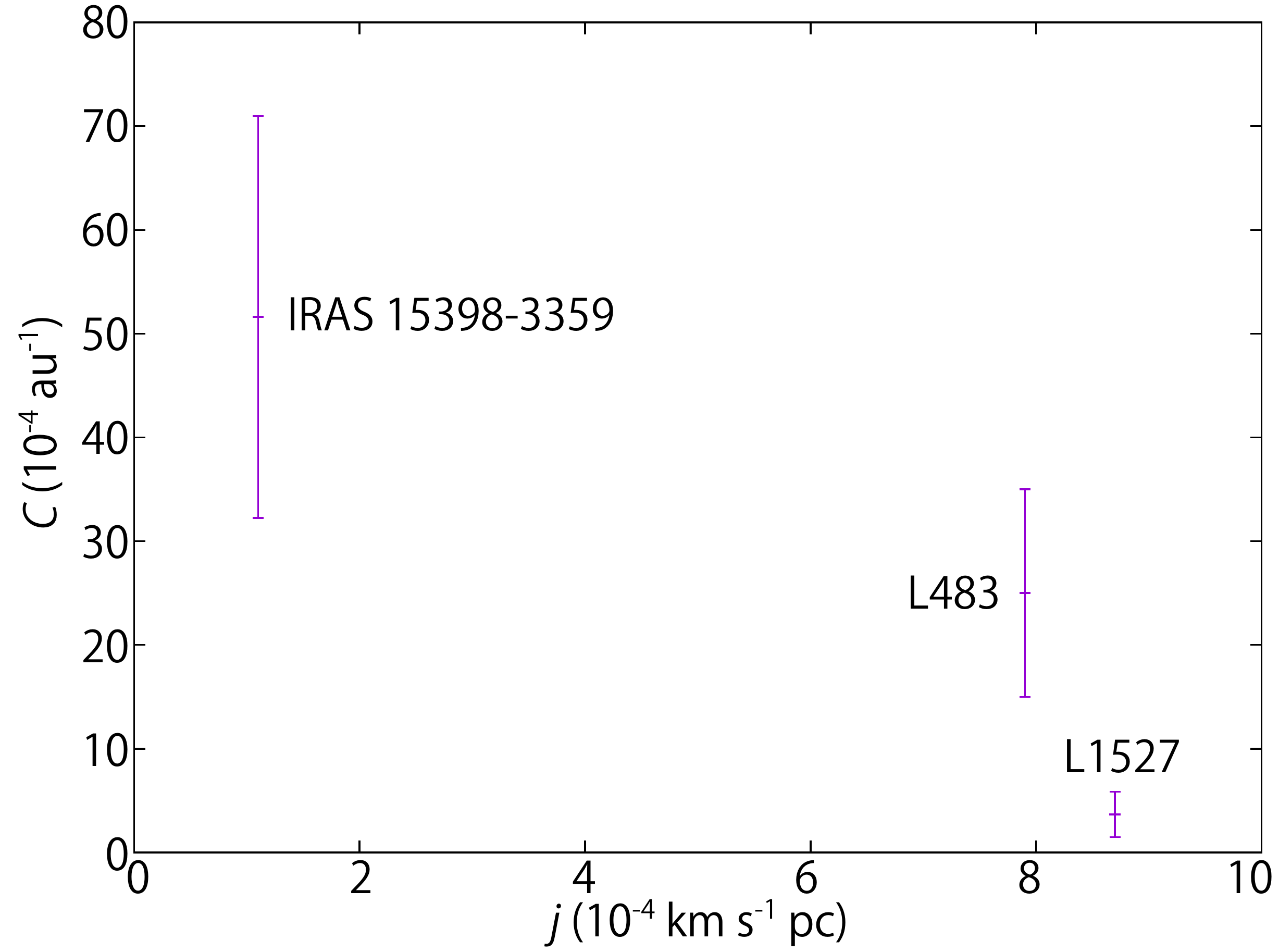}
	%\plotone{c-rcb_cerror.eps}
	\fi
	\caption{Plots of the curvature parameter ($C$) of the outflow against the \sam\ of the \ire. 
			%The value of the \sam\ for \irass\ is the upper limit. 
			%The correlation coefficient of this plot is $-0.90$, %$-0.8959031327451844$
			%where the uniform weights are applied to all the sources. 
			\label{fig:phys_CrCB}}
	\end{center}
\end{figure}

If the \cb\ is responsible for the launch of the low-velocity outflow, 
there would be some relations between the radius of the \cb\ and the outflow shape. 
%For instance, the outflow launched from the smaller radius of the \cb\ may be more collimated due to less centrifugal motion. 
To inspect its possibility, the outflow parameter $C$ is plotted against the \sam\ of the \ire\ ($j$; Figure \ref{fig:phys_CrCB}; Table \ref{tb:phys_ire}). 
{ 
It shows a hint of decrease of $C$ as increase of $j$. 
However, 
the data points are limited, and hence, 
} 
%a clear relation is not seen %between them 
%at the present stage. 
%due to the limited number of the data points. 
%there is a big difference in $C$ between L483 and L1527, having the similar \sam. 
%More samples are apparently needed to assess this possibility. 
{ this possible relation has to be followed up in future works.}

\section{Summary} \label{sec:summary}
We analyzed the kinematic structure of the outflow in the Class 0 low-mass protostellar source L483 
observed at a sub-arcsecond resolution with ALMA. 
The main results are summarized below: 

\begin{enumerate}
\item[(1)] 
The CCH and CS lines trace the bipolar outflow components 
extended at a 1000 au scale. 
On the other hand, the SO line, which is often regarded as the outflow tracer, 
does not trace the outflow components in this source. 

\item[(2)] 
Based on the distribution of the CS line and its kinematic structure, 
the geometrical configuration of the outflow is revealed. 
%The position angle of the outflow axis is 105\degr. 
The outflow blows almost on the \pos\ ($i \sim \parI$), 
where the northwestern and southeastern lobes are blue- and red-shifted, respectively. 

\item[(3)] 
The parabolic model of the outflow well reproduces the geometrical and velocity structures of the outflow traced by the CS line. 
The physical parameters for the model are evaluated as follows: 
$C =$ \parC, and $v_0 =$ \parV. 

\item[(4)]
The physical parameters for the parabolic outflow model are compared among seven protostellar sources. 
The parameters ($C$ and $v_0$) are further confirmed to decrease as the \dynT\ of the outflow, 
which means that the opening angle of the outflow increases. 
A possible relation between the outflow parameters and the \sam\ of the infalling envelope is suggested, 
but it has to be examined by observations of more sources. 

\item[(5)]
The rotation motion of the outflow is carefully inspected by using the outflow model 
taking account of the rotation. 
The PV diagrams seem to be better reproduced, 
if the \sam\ of the outflow cavity is 
{ a factor of a few of the \ire.} 
%twice that of the \ire. 
%When twice the \sam\ of 
%is assumed for the outflow, the PV diagrams seem to be better reproduced. 
Thus, a hint of the rotation motion is found, although it is marginal. 
\end{enumerate}

\acknowledgements
We use the ALMA data set ADS/JAO.ALMA\#2013.1.01102.S. 
ALMA is a partnership of the ESO (representing its member states), 
the NSF (USA) and NINS (Japan), together with the NRC (Canada) and the NSC and ASIAA (Taiwan), 
in cooperation with the Republic of Chile. 
The Joint ALMA Observatory is operated by the ESO, the AUI/NRAO and the NAOJ. 
We thank the ALMA staff for their support. 
We are also grateful for the financial support by KAKENHI 
%Y.O. acknowledges the JSPS fellowship. This study is supported by Grant-in-Aid from the Ministry of Education, Culture, Sports, Science, and Technologies of Japan 
%(21224002, 
(25400223, 25108005, and 15J01610). 
%N.S. and S.Y. acknowledge financial support by JSPS and MAEE under the Japan--France integrated action program (SAKURA: 25765VC).
Y.O., N.S., Y.W., and S.Y. acknowledge financial support by JSPS and MAEE under the Japan--France integrated action program.  
C.C. and B.L. thank CNRS for the support under the France--Japan action program. 

\appendix 

\section{\SAM} \label{sec:appendix_sam}

Here, we consider the \ire\ gas conserving the \sam\ \citep{Sakai_1527nature, Oya_15398}. 
In this case, the \sam\ ($j$) is represented in terms of the mass of the protostar ($M$) and the radius of the \cb\ (\rcb) as: 
\begin{align}
	j &= \sqrt{2GM r_{\rm CB}}, \label{eq:j_CB}
\end{align}
where $G$ denotes the gravitational constant. 
This is larger than the corresponding \sam\ for the Kepler motion at \rcb\ by a factor of $\sqrt{2}$: 
\begin{align}
	j &= \sqrt{GM r_{\rm CB}}. \label{eq:j_Kep}
\end{align}
\citet{Oya_483} investigated the kinematic structure of the envelope gas of L483 
with the aid of the \ire\ model, 
and evaluated the protostellar mass ($M$) of \parM\ and the radius of the \cb\ (\rcb) of \parRcbau. 
Here, they assumed the \ia\ ($i$) to be \parI\ (\incRem) derived from the outflow analysis in this paper. 
With the above values, we evaluate the \sam\ of the gas to be 
$7.9 \times 10^{-4}$ \kmps\ pc in the \ire\ by using Eq. (\ref{eq:j_CB}).

\section{Parabolic Outflow Model} \label{sec:appendix_model}
%In the analysis of the geometrical and velocity structures of the outflow, 
%the standard model of an outflow cavity \citep{Lee_outflow} is employed. 
We analyze the outflow structure by using the standard model reported by \citep{Lee_outflow}. 
Although this is just a morphological model, 
it has widely been applied to outflows in low-mass and high-mass star forming regions 
\citep[e.g.][]{Arce2013, Beuther2004, Lumbreras2014, Takahashi2012, Takahashi2013, Yeh2008, Zapata2014}. 
%(e.g. Arce et al. 2013; Beuther et al. 2004; Lumbreras \& Zapata 2014; Takahashi \& Ho 2012; Takahashi et al. 2013; Yeh et al. 2008; Zapata et al. 2014). 

This model assumes a parabolic shape of the outflow cavity. %that the outflow cavity has a parabolic shape. 
It also assumes that the outflow velocity linearly increases as an increasing distance to the protostar. 
Then, the shape and the velocity of the outflow cavity wall 
are represented as: 
\begin{align}
	z &= C R^2, \quad v_R = v_0 \frac{R}{R_0}, \quad v_z = v_0 \frac{z}{z_0}. 
\end{align}
Here, we define the $z$-axis along the outflow, 
where the origin is taken at the protostar position. 
We set the normalization constant $z_0$ to be 1 au. 
On the other hand, 
the radial size of the outflow cavity is denoted by $R$, 
where the normalization constant $R_0$ is set to be 1 au. 
We fit this model to the observed PV diagrams, 
and determine the two free parameters, $C$ and $v_0$. 

%where the $z$-axis is taken along the outflow axis ($z = 0$ at the protostar). 
%$R$ denotes the radial size of the outflow cavity perpendicular to the $z$-axis. 
%$R_0$ and $z_0$ are normalization constants, both equals to 1 au. 
%Remaining constants, $C$ and $v_0$ are free parameters. 
%Thus, the outflow cavity wall has a parabolic shape in this model, 
%and it is linearly accelerated as the distance from the protostar along the outflow axis ($z$) 
%and that from the outflow axis ($R$). 

The outflow parameters for \irasVLA, RNO 91, L1448C, and HH 46/47 are originally reported in the unit of arcsecond
\citep{Lee_outflow, Hirano2010, Arce2013}: 
\begin{align}
	C_{\rm as} &= C D, \quad v_{\rm as} = v_0  D. \label{eq:phys_outflow_as}
\end{align}
Here, $D$ denotes the source distance.  
Hence, the coefficients of proportionality are converted to the values in the unit of au 
in Table \ref{tb:phys_outflow}. 

{ 
The opening angle of the outflow ($\theta$) can be defined for a fixed distance of $a$. 
At the distance of $a$ along the outflow axis from the protostar (i.e. the length of the outflow), 
the radial size of the outflow is $\sqrt{a/C}$ in radius. 
Thus, the opening angle is expressed as: 
\begin{align}
	\tan \frac{\theta}{2} &= \frac{\sqrt{a / C}}{a} \nonumber \\ 
	&=	\frac{1}{\sqrt{a C}}. \label{eq:openangle}
\end{align}
For instance, 
the opening angle of the outflow model for L483 is calculated to be 
42\degr, 32\degr, and 24\degr, 
for $a$ of 500, 1000, and 2000 au, respectively. 
%180./pi*atan(1./sqrt(z0*C))
It should be noted that the opening angle decreases as increasing $a$, 
and thus the $a$ value should be fixed 
when comparing the opening angle among sources. 
Also, it should be stressed that the obtained opening angle is 
no longer under an influence of the \ia\ effect. 
}

%%%%%%%%%%%%%%%%%%%%%%%%%%%%%%%

%%%%%%%%%%%%%%%%%%%%%%%%%%%%%%%
\begin{landscape}
\begin{table}
	\begin{center}
	\caption{Parameters of the Observed Lines\tablenotemark{a} 
			\label{tb:lines}}
	\steptbnote
	\begin{tabular}{llccccc}
		\hline \hline 
		Molecule & Transition & Frequency (GHz) & $E_u$ (K) & $S\mu^2$ (Debye$^2$) & $A_{ij}$ ($s^{-1}$) & Synthesized Beam \\ \hline
		CS & \cs & 244.9355565 & 35.3 & 19 & 2.98 $\times 10^{-4}$ & $0\farcs51 \times 0\farcs46$ (P.A. $2.\!\!\degr76$) \\ %(P.A. $-177.\!\!\degr24$) \\ % Elow = 16.3411 cm-1 \\ 
		& & & & & & $0\farcs99 \times 0\farcs95$ (P.A. $-71.\!\!\degr95$)\tablenotemark{b} \\ %(P.A. $-177.\!\!\degr24$) \\ % Elow = 16.3411 cm-1 \\ 
		%\FA\tbnotemark & \fa & 247.390719 & 78.1 & 156 & 1.10 $\times 10^{-3}$ & $0\farcs56 \times 0\farcs49$ (P.A. 16.\!\!\degr44) \\ % Elow = 46.0459 cm-1 \\ 
		%\MF\tablenotemark{c} & \mf & 249.0474280 & 141.6 & 50 & 1.46 $\times 10^{-3}$ & $0\farcs52 \times 0\farcs45$ (P.A. $-177.\!\!\degr18$) \\ % Elow = 90.0826 cm-1 \\
		%SiO\tablenotemark{c} & \sio & 260.5180090 & 43.8 & 58 & 9.12 $\times 10^{-4}$ & $0\farcs46 \times 0\farcs42$ (P.A. $-177.\!\!\degr78$) \\ % Elow = 21.7261 cm-1 \\
		%\AAL\tablenotemark{c} & \aal & 260.5440195 & 96.4 & 82 & 6.25 $\times 10^{-4}$ & $0\farcs46 \times 0\farcs42$ (P.A. $-177.\!\!\degr81$) \\ 
		SO & \so & 261.8437210 & 47.6 & 16 & 2.28 $\times 10^{-4}$ & $0\farcs46 \times 0\farcs42$ (P.A. 3.\!\!\degr09) \\ % Elow = 24.3157 cm-1 \\ 
		CCH\tbnotemark & $N=3-2, J=7/2-5/2$ & & & & & \\ 
		 & $F=4-3$ & 262.0042600 & 25.1 & 2.3 & 5.32 $\times 10^{-5}$ & $0\farcs98 \times 0\farcs92$ (P.A. $-75.\!\!\degr93$) \\ %(P.A. $-78.\!\!\degr06$) \\ % Elow = 8.7402, 8.7393 cm-1 \\ 
		 & $F=3-2$ & 262.0064820 & 25.1 & 1.7 & 5.12 $\times 10^{-5}$ & $0\farcs98 \times 0\farcs92$ (P.A. $-75.\!\!\degr93$) \\ %(P.A. $-78.\!\!\degr06$) \\ 
		%t-HCOOH\tablenotemark{c} & \hcooh & 262.103481 & 82.8 & 24 & 2.03 $\times 10^{-4}$ & $0\farcs46 \times 0\farcs42$ (P.A. 2.\!\!\degr98) \\ 
		%\DEE\tablenotemark{c} & \dee & 262.393513 & 118.0 & 148 & 7.18 $\times 10^{-5}$ & $0\farcs46 \times 0\farcs41$ (P.A. 1.\!\!\degr69) \\ % Elow = 73.2646 cm-1 \\
		%HNCO\tablenotemark{c} & \hnco & 263.7486250 & 82.3 & 30 & 2.56 $\times 10^{-4}$ & $0\farcs48 \times 0\farcs41$ (P.A. 3.\!\!\degr87) \\ % Elow = 48.3904 cm-1 \\ 
		\hline 
	\end{tabular}
	\end{center}
	\resettbnote
	\tbnotetext{Taken from CDMS \citep{Muller_CDMS}. } %(M\"{u}ller et al. 2005) 
					%and JPL \citep{Pickett_JPL}. } %(Pickett et al. 1998). }
	%\tbnotetext{Nuclear spin degeneracy is not included. } 
	%\tbnotetext{4 channel bind. The spectral profile of the t-HCOOH line (Figure \ref{fig:spectra}) is obtained with binding 16 channels. } 
	\tbnotetext{An outer taper of 1\arcsec\ is applied to improve the S/N ratio. } 
	\resettbnote
\end{table}
\end{landscape}

\begin{table}
	\begin{center}
	\caption{Physical Parameters of L483 \label{tb:params_483}}
	\resettbnote
	\begin{tabular}{lrc}
	\hline \hline 
	\multicolumn{3}{c}{Outflow} \\ \hline
	\multicolumn{2}{l}{Inclination Angle\tbnotemark} & \parI \\ 
	& & (The NW lobe is blue-shifted.) \\ 
	\multicolumn{2}{l}{Position Angle} & 105\degr \\ 
	\multicolumn{2}{l}{Dynamic Timescale (\tdyn)\tbnotemark} & $3 \times 10^3$ yr \\ 
	\multicolumn{2}{l}{Outflow Mass Loss Rate (\Mout)\tbnotemark} & $(5.4 - 14) \times 10^{-6}$ \UnitMrate \\
	\hline
	\multicolumn{3}{c}{Protostar/Envelope} \\ \hline 
	\multicolumn{2}{l}{Age\tbnotemark} & $< 4.2 \times 10^3$ yr \\
	& & (With an uncertainty of a factor of 3) \\  
	\multicolumn{2}{l}{Central Mass (\Mstar)\tbnotemark} & 0.15 \Msun \\ 
	\multicolumn{2}{l}{Bolometric Luminosity (\Lbol)\tbnotemark} & 13 \Lsun \\ 
	\multicolumn{2}{l}{Bolometric Temperature (\Tbol)\tablenotemark{d}} & $< 56$ K \\ 
	Accretion Rate (\Macc)\tbnotemark & $(M_{\rm star} / t_{\rm dyn})$ & $5 \times 10^{-5}$ \UnitMrate \\
	& $(L R_{\rm star} / G M_{\rm star})$ & $6.9 \times 10^{-6}$ \UnitMrate \\ 
	\multicolumn{2}{l}{\SAM\ ($j$)\tbnotemark} & $7.9^{+4}_{-3} \times 10^{-4}$ \kmps\ pc \\ 
	\hline 
	\end{tabular}
	\end{center}
	\resettbnote
	\tbnotetext{\incRem. } 
	\tbnotetext{See Section \ref{sec:outflow_model}. }
	\tbnotetext{Taken from \citet{Yildiz_CO-highJ}. }
	\tbnotetext{Taken from \citet{Fuller1995}. } 
	\tbnotetext{Taken from \citet{Oya_483}. } 
	\tbnotetext{Taken from \citet{Shirley2000}. } 
	\tbnotetext{See Section \ref{sec:env-outflow_mass}. } 
	\tbnotetext{See Appendix \ref{sec:appendix_sam}. } 
	\resettbnote
\end{table}

\begin{landscape}
\begin{table}
	\begin{center}
	\caption{Phsyical Parameters of Outflows \label{tb:phys_outflow}}
		\resettbnote
		\begin{tabular}{lccccccc}
			\hline \hline
			Source & Distance & $T_{\rm bol}$\tbnotemark & \tdyn & Inclination Angle\tbnotemark & $C$ & $v_0$ \\ 
			%& (pc) & ($10^3$ yr) & ($\degr$) & ($10^{-4}$ au$^{-1}$) & ($10^{-4}$ km s$^{-1}$ au$^{-1}$) \\ %& (km s$^{-1}$ au$^{-\frac{1}{2}}$) \\ 
			& (pc) & (K) & ($10^3$ yr) & ($\degr$) & ($10^{-4}$ au$^{-1}$) & ($10^{-4}$ km s$^{-1}$) \\ %& (km s$^{-1}$ au$^{-\frac{1}{2}}$) \\ 
			\hline 
			L483 & 200 & 52 & 3 & 80 $\pm$ 5 & $25 \pm 10$ & $15 \pm 5$ \\ 
			\irass \tbnotemark & 155 & 44 & 0.9 (northeast), 1.8 (southwest)\tbnotemark & 70 $\pm$ 10 & 52 $\pm$ 19 & 25 $\pm$ 2 \\ %& 0.034 \\ 
			L1527\tbnotemark & 137 & 67 & 20.6 (east), 6.5 (west)\tablenotemark{d} & 85 $\pm$ 10 & 3.6 $\pm$ 2.2 & 7.3 $\pm$ 1.5 \\ %& 0.050 ($\pm$ 0.019?) \\ % offset=0."62, C = 0.05\pm0.03, v0 = 0.1
			\irasVLA \tbnotemark & 460 & & $\sim$7& 71 $\pm$ 3 & 4.8 $\pm$ 1.1 & 7.0 $\pm$ 1.1 \\ %& 0.032 $\pm$ 0.005 \\ 
			RNO 91\tablenotemark{f} & 160 & 715 & $\sim$20 & 70 $\pm$ 4 & 1.3 $\pm$ 0.25 & 2.4 $\pm$ 0.6 \\ %& 0.021 $\pm$ 0.009 \\ 
			L1448C\tbnotemark & 250 & 54 & $\sim 0.24$& 69 & 24, 32 & 200 \\ %& 0.41, 0.35 \\ 
			HH 46/47\tbnotemark & 450 & 111.0 & 9 & 61 $\pm$ 1 & 6.7 $\pm$ 1.1 & 51 $\pm$ 4 \\ %& 0.20 $\pm$ 0.02 \\ 
			\hline 
		\end{tabular}
		\vspace*{-30pt}
	\end{center}
	\resettbnote
	\tbnotetext{Bolometric temperatures are taken 
				from \citet{Shirley2000} for L483 and L1448C, 
				from \citet{Green2013} for L1527, 
				from \citet{Jorgensen_15398} for \irass, 
				from \citet{Yang2018} for HH46/47, 
				and from \citet{Chen_Tbol} for RNO91. 
				}
	\tbnotetext{\incRem. } 
	\tbnotetext{Determined from the \FAD\ (\fad) emission \citep{Oya_15398}. } 
	\tbnotetext{Determined from the CO ($J=3-2$) emission \citep{Yildiz_CO-highJ}. } %(Y\i ld\i z et al. 2015). }
	\tbnotetext{Determined from the CS (\cs) emission \citep{Oya_1527}. } 
	\tbnotetext{Determined from the CO ($J=1-0$) emission \citep{Lee_outflow}. 
			\irasVLA\ is a Class 0/I source in the Orion dark cloud L1617, 
			while RNO 91 is a Class II/III source in the L43 molecular cloud. } %(Lee et al. 2000). }
	\tbnotetext{Determined from the CO ($J=3-2$) emission \citep{Hirano2010}. 
			L1448C (L1448 mm) is a Class 0 source in Perseus. } %(Hirano et al. 2010). }
	\tbnotetext{Determined from the CO ($J=1-0$) emission \citep{Arce2013}. 
			HH 46/47 molecular outflow is on the outskirt of the Gum Nebula. } %(Arce et al. 2013). }
	\resettbnote
\end{table}
\end{landscape}

%\begin{landscape}
\begin{table}
	\resettbnote
	\begin{center}
	\caption{Physical Parameters of Envelopes \label{tb:phys_ire}}
	\begin{tabular}{lccccccc}
	\hline \hline 
	\multirow{2}{*}{Source Name} & Evolutionary & Inclination & Protostellar & Radius of & $j$ \\ 
	& Stage & Angle (\degr)\tbnotemark & Mass (\Msun) & the CB\tbnotemark (au) & ($10^{-4}$ \kmps\ pc)\tbnotemark \\ \hline 
	L483  & Class 0 & 80 & $0.15 \pm 0.05$ & $100^{+100}_{-70}$ & $7.9^{+4}_{-3}$ \\ 
	%\irass \tbnotemark & Class 0/I & 70 & $0.02 \pm 0.02$ & $<$30 & $<$1.6 \\ 
	\irass \tbnotemark & Class 0/I & 70 & $0.007^{+0.004}_{-0.003}$ & 40 & $1.1 \pm 0.3$ \\ 
	L1527\tbnotemark & Class 0/I & 85 & $0.18 \pm 0.05$ & $100 \pm 20$ & $8.7 \pm 1$ \\ 
	\hline 
	\end{tabular}
	\end{center}
	\resettbnote
	\tbnotetext{\incRem. These values are derived with the aid of the parabolic outflow model (see Appendix \ref{sec:appendix_model}). }  
	\tbnotetext{Centrifugal barrier. } 
	\tbnotetext{Specific \am\ of the envelope gas 
				derived from the protostellar mass and the radius of the \cb\ (see Appendix \ref{sec:appendix_sam}). } 
	\tbnotetext{Taken from \citet{Oya_15398} and \citet{Okoda_15398}. } 
	\tbnotetext{Taken from \citet{Oya_1527}. } 
	\resettbnote
\end{table}
%\end{landscape}

\begin{landscape}
\begin{figure}
	\begin{center}
	\iffigure
	%\epsscale{1.0}
	%\plotone{fig_moment_extended.eps}
	\includegraphics[bb = 900 0 2000 1400, scale = 0.26]{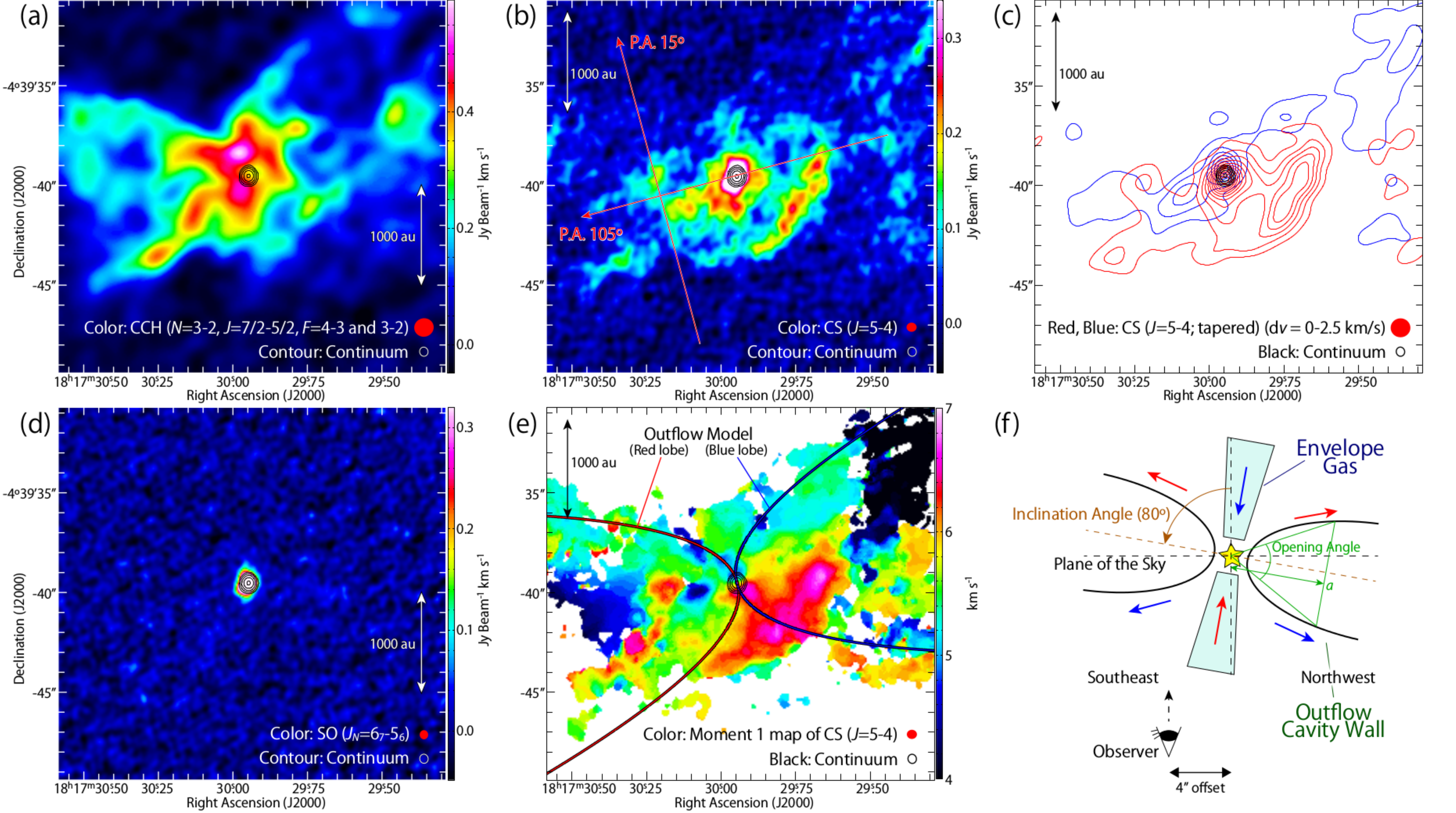}
	\fi
	\end{center}
\end{figure}
\end{landscape}
\begin{landscape}
\begin{figure}
	\begin{center}
	\caption{Integrated intensity maps of CCH (\cch; a), CS (\cs; b, c), and SO (\so; d). 
			%velocity range: -2.5 ~ 13.5 for CCH and CS (b), 0 ~ +-2.5 for CS (c), -3.5 ~ 14.5 for SO
			{\bf 
			An outer taper of 1\arcsec\ is applied for the CS data in panel (c). 
			} 
			The velocity-shift range for integration is $\pm8$ \kmps\ for CCH and CS (b), 
			and $\pm9$ \kmps\ for SO \citep[$v_{\rm sys} = 5.5$ \kmps;][]{Hirota_carbonchain}. 
			The CCH emission has the contribution from both the two hyperfine components (see Table \ref{tb:lines}). 
			In panel (c) (CS), the velocity-shift range for integration is 
			$(0 - +2.5)$ and $(-2.5 - 0)$ \kmps\ for the red and blue contours, respectively. 
			The outflow component is clearly seen in the maps for the narrower integration rarnges. 
			{\bf 
			The contour levels for the red and blue contours are every 5$\sigma$ from 3$\sigma$, %every 10$\sigma$ from 3$\sigma$, 
			where the rms noise level is 20 mJy beam\inv\ \kmps. %6 mJy beam\inv\ \kmps. 
			}
			The black contours represent the 1.2 mm continuum map, 
			where the contour levels are 10, 20, 40, 80, and 160$\sigma$, where the rms noise level is 0.13 mJy beam\inv. 
			The outflow axis is along the red arrow with a P.A. of 105\degr\ in panel (b). 
			The PV diagram in Figure \ref{fig:PV_CSoutflow}(b) is prepared along the red arrow with a P.A. of 15\degr\ in panel (b), 
			which is centered at the position with an offset of 4\arcsec\ to the southeast from the continuum peak along a P.A. of 105\degr. 
			{ 
			Panel (e) shows the velocity field (moment 1) map of CS 
			with the velocity range for integration of $(-2.5 - +2.5)$ \kmps, 
			on which the parabolic outflow model is superposed (see Section \ref{sec:outflow_model}). 
			}
			Panel (f) shows a schematic illustration of the outflow, 
			which blows almost on the \pos. 
			\label{fig:mom0_extended}}
	\end{center}
\end{figure}
\end{landscape}

\begin{landscape}
\begin{figure}
	\begin{center}
	\iffigure
	%\epsscale{1.0}
	%\plotone{PV.CS.centRa18h17m29s947Dec-04d39m39s52.PA105deg.SE4asOffset.color_-3-20sigma.contourPreset.outflow.D200.I-10G0.5Vas0.3VP1_lee_k7A11_norotation-01.eps}
		\includegraphics[bb = 0 0 1400 600, scale = 0.41]{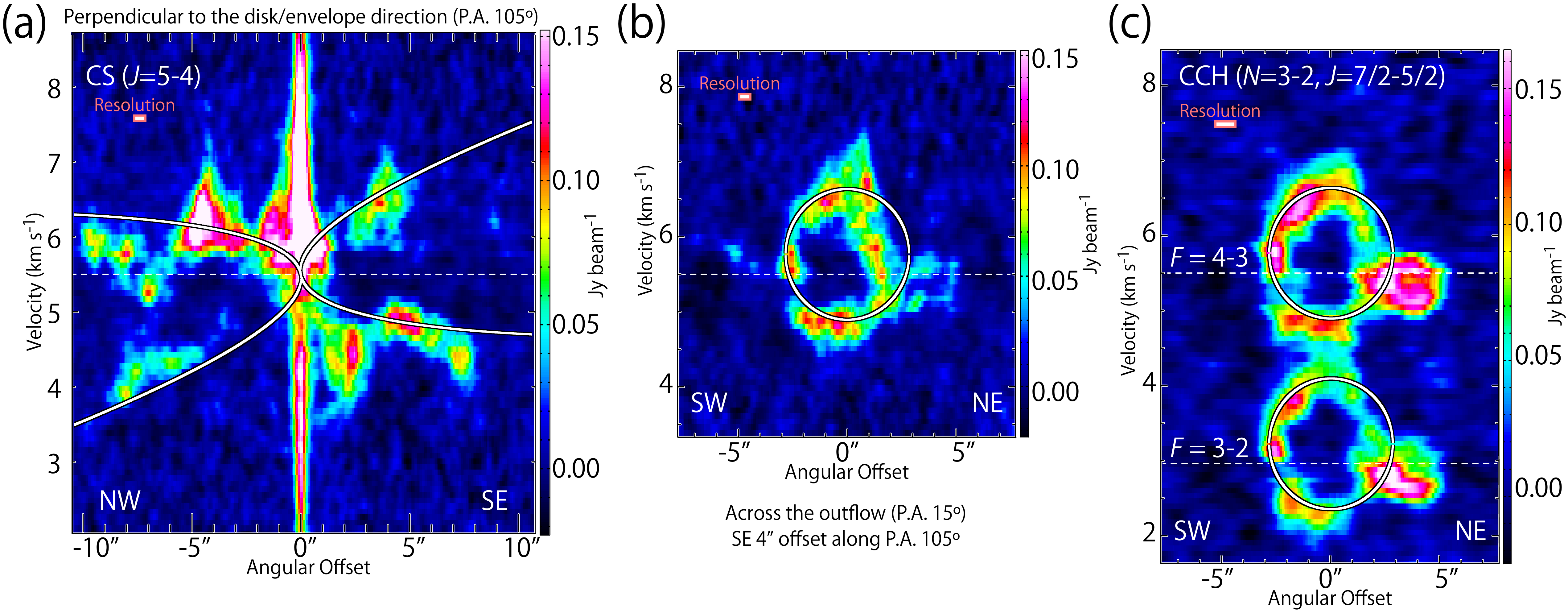}
	\fi
	\caption{Position-velocity diagrams of CS (a, b) and { CCH (c)} along the outflow axis (a; P.A. 105\degr) 
			and the line across the outflow (b, c; P.A. 15\degr) shown in Figure \ref{fig:mom0_extended}(b). 
			The position axis in panels (b, c) is centered at the distance of 4\arcsec\ from the protostellar position toward the southeastern direction. 
			%The contour levels are every 5$\sigma$, where the rms noise level is 7.6 mJy beam\inv. 
			{ White lines} represent the results of the parabolic outflow model, 
			where the physical parameters are as follows; 
			$i$ = \parI, $C$ = \parC, and $v_0$ = \parV. 
			%The contour levels for the outflow model are 10\%, 50\%, and 90\%\ to the peak intensity. 
			%We consider the paraboloid convolved by the spatial and velocity resolutions (see Section \ref{sec:outflow_model}). 
			{ 
			In panel (c), the outflow model is prepared for each component of the two hyperfine components, 
			whose separation is 2.54 \kmps. 
			}
			\label{fig:PV_CSoutflow}}
	\end{center}
\end{figure}
\end{landscape}

\begin{figure}
	\begin{center}
	\iffigure
	%\epsscale{1.0}
	%\plotone{PV.CS.5-4.selfcal.vwid41.manucleaned.robust0.5.niter3000.threshold6mJy.notaper.SW0-10asNW0-10asfromContPeakalongPA105deg.PA015.0deg.outflow.color-3-15sigma.contourPreset.SE_D200.I-10G0.5Vas0.3VP1_lee_k7A11_nomodel.eps} 
	\includegraphics[bb = 20 0 2000 1700, scale = 0.27]{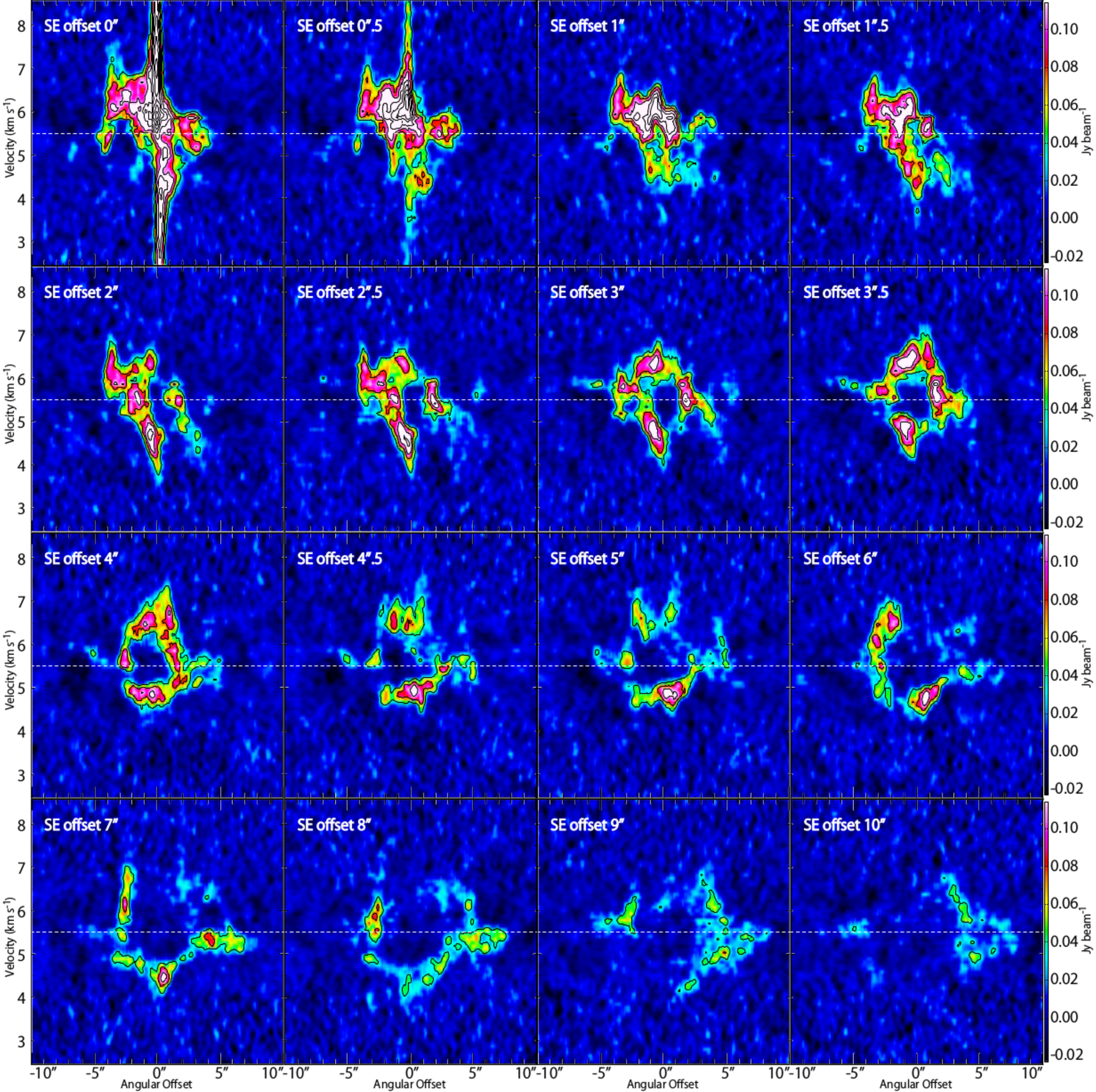}
	\fi
	\caption{Position-velocity diagrams of CS (\cs) across the outflow axis. 
			The position axes are along the position angle of 15\degr, 
			which are perpendicular to the outflow axis (P.A. 105\degr). 
			The origin of the position axes is on the outflow axis with offsets of 
			the distance of ($0\arcsec-10\arcsec$) from the protostellar position 
			toward the southeastern direction. 
			The color map in the panel labeled as `SE offset 4\arcsec' is the same as that in Figure \ref{fig:PV_CSoutflow}(b). 
			The contour levels for CS are every 5$\sigma$, where the rms noise level is 7.6 mJy beam\inv. 
			\label{fig:outflow_SEoffset-nomodel}}
	\end{center}
\end{figure}

\clearpage

\clearpage
\begin{figure}
	\begin{center}
	\iffigure
	%\epsscale{1.0}
	%\plotone{PV.CS.5-4.selfcal.vwid41.manucleaned.robust0.5.niter3000.threshold6mJy.notaper.SW0-10asNW0-10asfromContPeakalongPA105deg.PA015.0deg.outflow.color-3-15sigma.contourPreset.SE_D200.I-10G0.5Vas0.3VP1_lee_k7A11_rotx0.1.2.4.eps} 
	\includegraphics[bb = 20 0 2000 1700, scale = 0.27]{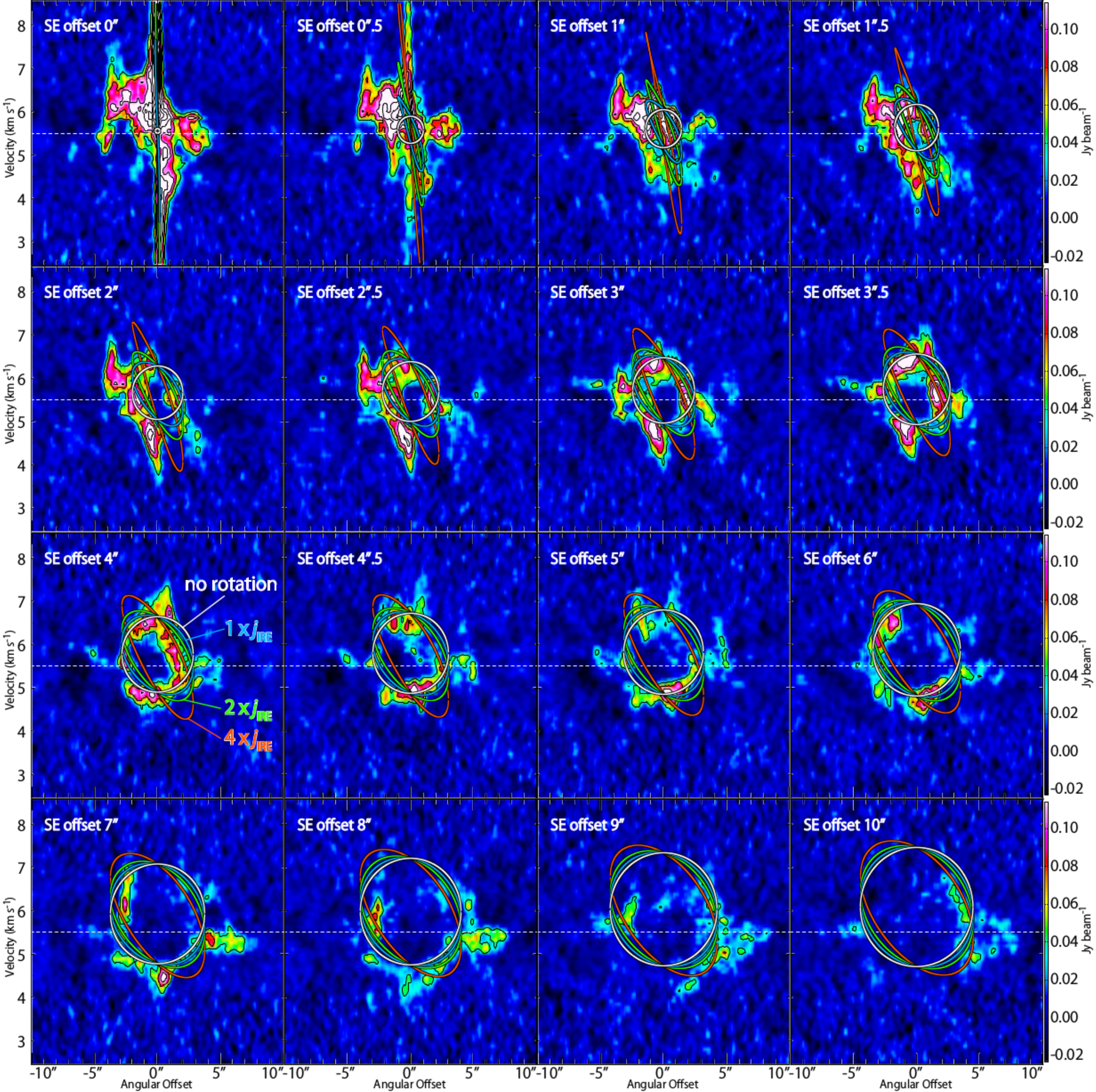}
	\fi
	\caption{Position-velocity diagrams of CS (\cs) across the outflow axis. 
			The color maps and the black contours are the same as those in Figure \ref{fig:outflow_SEoffset-nomodel}. 
			Elliptic lines represent the results of the outflow model, 
			where the physical parameters are as follows; 
			$i$ = \parI, $C$ = \parC, and $v_0$ = \parV. 
			The \sam of the gas in the outflow is assumed to be 
			$0$, $7.9 \times 10^{-4}$, $15.8 \times 10^{-4}$, and $31.6 \times 10^{-4}$ \kmps\ pc 
			for the white, blue, green, and red lines, respectively. 
			The values for the latter three models correspond to once, twice, and four times 
			of the \sam\ of the \ire\ ($j_{\rm IRE} = 7.9 \times 10^{-4}$ \kmps\ pc). 
			The \sam\ of the gas is assumed to be conserved in each outflow model. 
			%The contour levels for the outflow model are 10\%, 50\%, and 90\%\ to the peak intensity in each panel. 
			\label{fig:outflow_SEoffset-withmodel}}
	\end{center}
\end{figure}

\clearpage
\begin{figure}
	\begin{center}
	\iffigure
	%\epsscale{1.0}
	%\plotone{PV.CS.5-4.selfcal.vwid41.manucleaned.robust0.5.niter3000.threshold6mJy.notaper.SW0-10asNW0-10asfromContPeakalongPA105deg.PA015.0deg.outflow.color-3-15sigma.contourPreset.NW_D200.I-10G0.5Vas0.3VP1_lee_k7A11_rotx0.1.2.4.eps} 
	\includegraphics[bb = 20 0 2000 1700, scale = 0.27]{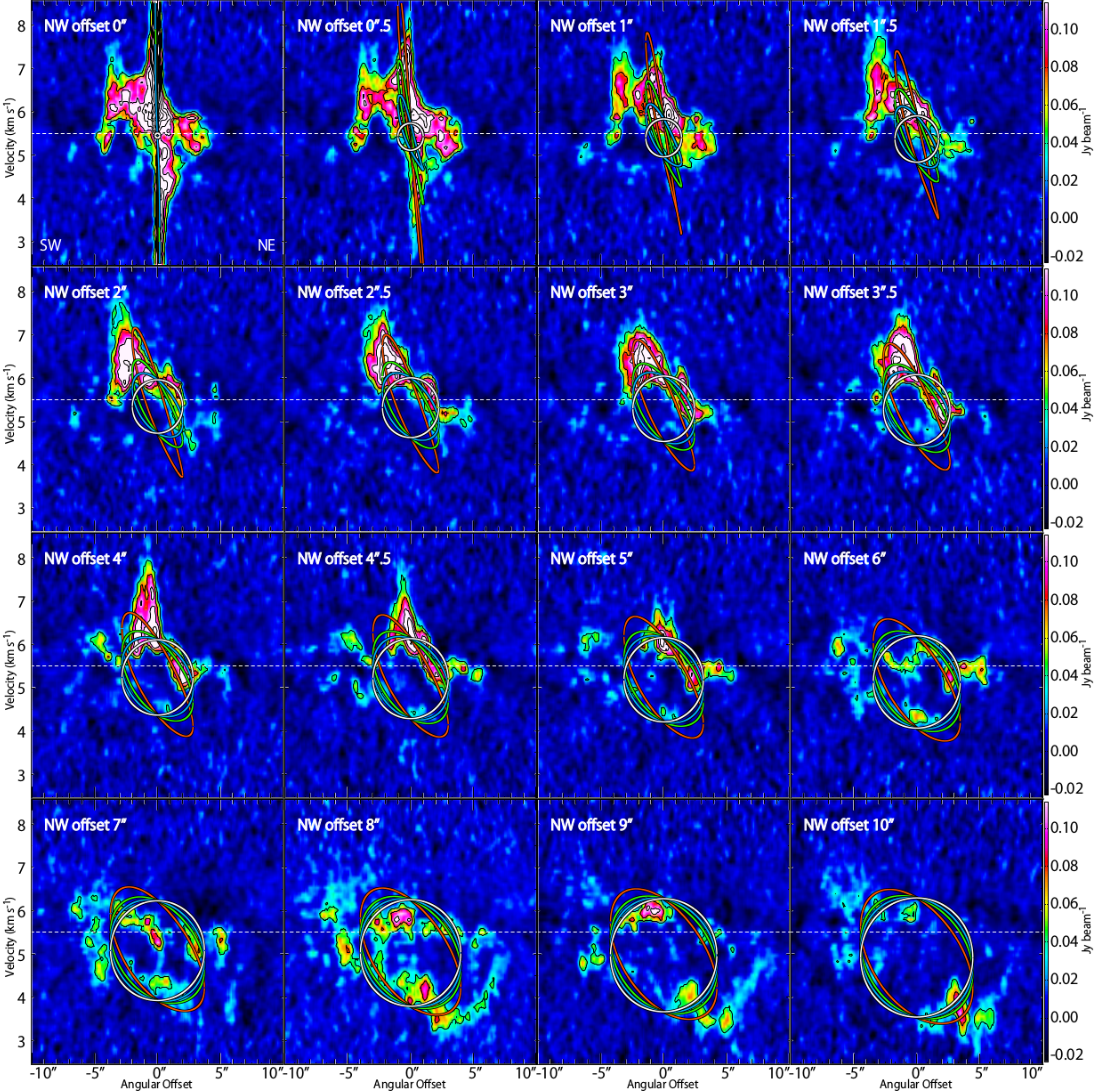}
	\fi
	\caption{Position-velocity diagrams of CS (\cs) across the outflow axis. 
			The color maps and the black contours are the same as those in Figure \ref{fig:outflow_NWoffset-nomodel}. 
			Elliptic lines represent the results of the same outflow models shown in Figure \ref{fig:outflow_SEoffset-withmodel}. 
			\label{fig:outflow_NWoffset-withmodel}}
	\end{center}
\end{figure}

\clearpage

\clearpage
\begin{figure}
	\begin{center}
	\iffigure
	%\epsscale{1.0}
	%\plotone{V0vs_tdyn-Tbol.eps}
	\includegraphics[bb = 0 0 1800 100, scale = 0.325]{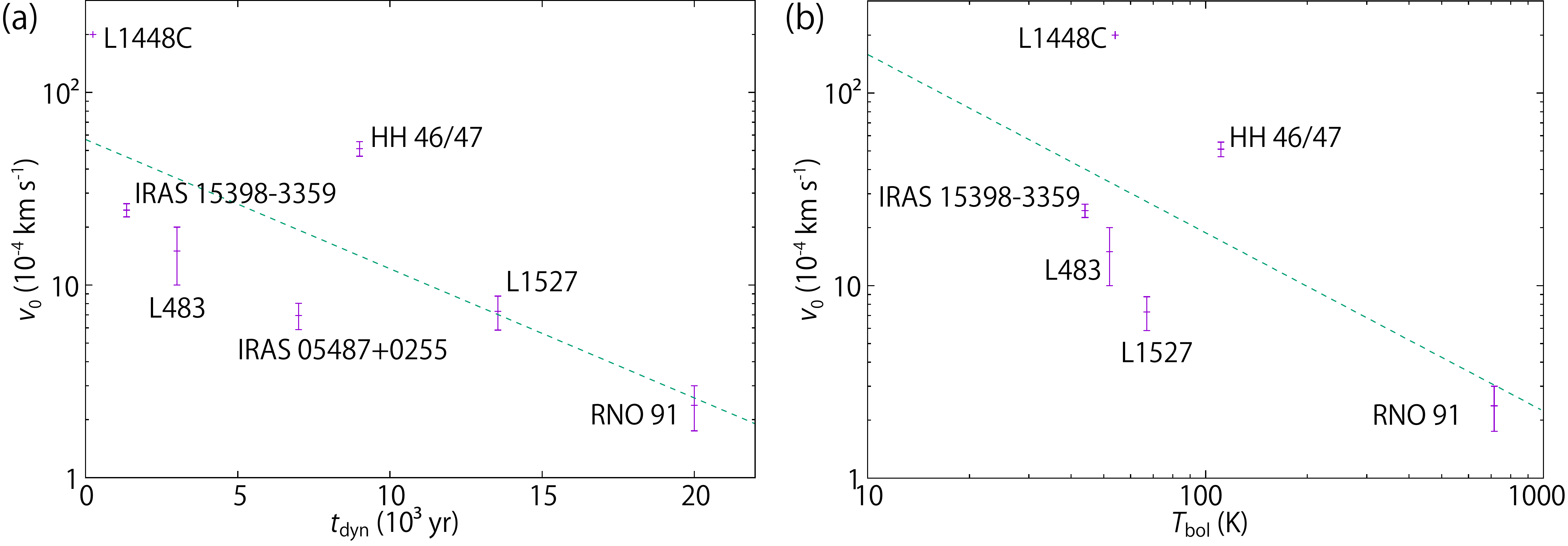}
	\fi
	\caption{Relation between the velocity parameter ($v_0$) of an outflow, { its \dynT\ (\tdyn; a), and its \bolT\ (\Tbol; b).}  
			The \dynT s for \irass\ and L1527 are the averaged value of the blue and red lobes. 
			The green dashed lines represent the best-fit functions: 
			$\log v_0 = (-0.15 \pm 0.06) \times (t_{\rm dyn} \times 10^{-3}) + (-5.2 \pm 0.6)$ %L483 tdyn = 3x10^3 yr
			%$\log v_0 = (-0.17 \pm 0.06) \times (t_{\rm dyn} \times 10^{-3}) + (-4.9 \pm 0.7)$, %L483 tdyn = 2x10^3 yr
			{ 
			and 
			$\log v_0 = (-0.9 \pm 0.6) \times \log$ \Tbol $ + (-2 \pm 3)$ 
			for (a) and (b), respectively. 
			}
			The correlation coefficients 
			{ are $-0.75$ %$-0.74$, %$-0.80$, 
			and $-0.63$ 
			for (a) and (b), respectively.} 
			where the uniform weights are applied to all the sources. 
			\label{fig:phys_v0Dt}}
	\end{center}
\end{figure}

\clearpage

\clearpage

%{ (Machida \& Hosokawa (2013) と比較 \rarrow\ theoretical modelのB, j, etc. に制限を与えること)}

\end{document}